\documentclass[superscriptaddress,longbibliography,aps,pra,preprintnumbers]{revtex4-2}
\usepackage[urlcolor=blue,colorlinks=true,citecolor=blue,linkcolor=blue,pdfstartview={FitH},bookmarks=false]{hyperref}
\usepackage{graphicx,comment,float}
\usepackage{xcolor,soul}
\usepackage{dcolumn}
\usepackage{bm}
\usepackage{amsmath,amssymb}
\usepackage{mathrsfs} 
\usepackage[normalem]{ulem}
\usepackage{url}
\usepackage{subfig}
\usepackage{lineno}

\usepackage{helvet}

\newcommand{\KS}[1]{\textcolor{blue}{#1}}
\begin{document}
\title{Nature of field-induced transitions and hysteretic magnetoresistance \\ in the noncollinear antiferromagnet EuIn$_2$As$_2$} 


\author{Karan Singh}
\email[e-mail: ]{k.singh@intibs.pl}
\affiliation{Institute of Low Temperature and Structure Research, Polish Academy of Sciences, Okólna 2, 50-422 Wrocław, Poland}

\author{Jan Skolimowski}
\affiliation{International Research Centre MagTop, Institute of Physics,
Polish Academy of Sciences, Aleja Lotników 32/46, PL-02668 Warsaw, Poland}

\author{Giuseppe Cuono}
\affiliation{Consiglio Nazionale delle Ricerche (CNR-SPIN), Unit\'a di Ricerca presso Terzi c/o Universit\'a “G. D’Annunzio”, 66100 Chieti, Italy}

\author{Raghottam M. Sattigeri}
\affiliation{Physics Department, Università degli Studi di Milano, Via Celoria 16,
20133 Milan, Italy}

\author{Andrzej Ptok}
\affiliation{Institute of Nuclear Physics, Polish Academy of Sciences, W. E. Radzikowskiego 152, PL-31342 Kraków, Poland}

\author{Orest Pavlosiuk}
\affiliation{Institute of Low Temperature and Structure Research, Polish Academy of Sciences, Okólna 2, 50-422 Wrocław, Poland}

\author{Tetiana Romanova}
\affiliation{Institute of Low Temperature and Structure Research, Polish Academy of Sciences, Okólna 2, 50-422 Wrocław, Poland}

\author{Tomasz Toliński}
\affiliation{Institute of Molecular Physics, Polish Academy of Sciences,
M. Smoluchowskiego 17, 60-179 Poznań, Poland}

\author{Piotr Wiśniewski}
\affiliation{Institute of Low Temperature and Structure Research, Polish Academy of Sciences, Okólna 2, 50-422 Wrocław, Poland}

\author{Carmine Autieri}
\email[e-mail: ]{autieri@magtop.ifpan.edu.pl}
\affiliation{International Research Centre MagTop, Institute of Physics,
Polish Academy of Sciences, Aleja Lotników 32/46, PL-02668 Warsaw, Poland}

\author{Dariusz Kaczorowski}
\email[e-mail: ]{d.kaczorowski@intibs.pl}
\affiliation{Institute of Low Temperature and Structure Research, Polish Academy of Sciences, Okólna 2, 50-422 Wrocław, Poland}

\date{\today}
             
\begin{abstract}

We examine the magnetic and electrical transport properties of the hexagonal EuIn$_2$As$_2$ compound, combining experimental and theoretical results. This compound is predicted to be an axion-insulator from an electronic point of view and an altermagnet while in the collinear magnetic phase. However, experiments indicate that the Fermi level lies within the valence band rather than in the topological gap, potentially leading to the dominance of magnetic properties. Our detailed studies on magnetization and electrical transport support the presence of a broken-helix antiferromagnetic state, which was previously identified by X-ray and neutron diffraction experiments. 
Notably, we observed within that state a field-induced metamagnetic transition marked by a large hysteresis in magnetoresistance, which turns into a sharp upturn for the magnetic field tilted by 15$^\circ$ from the \textit{c}-axis of the crystal.
Combined with theoretical calculations, it is explained that the application of a magnetic field changes the low-resistivity antiferromagnetic domain walls to the high-resistivity domain walls due to the reduction in the Fermi surface sheets interaction area in the domain walls, originating from \textit{p}-orbitals of As. EuIn$_2$As$_2$, therefore, presents a new case study that broadens the understanding of complex magnetic structures and their influence on electrical transport.

\end{abstract}

\maketitle

\begin{center}
\noindent\textbf{I. INTRODUCTION}
\end{center}

In recent years, considerable attention has been drawn to the ternary europium-based antiferromagnet EuIn$_2$As$_2$  due to exotic physics originating from the interplay of topological and magnetic properties \cite{Xu2019, Riberolles2021,Soh2023, Donoway2024}.
The compound exhibits two subsequent antiferromagnetic (AFM) transitions at $T_{N1}=$17.6 K and $T_{N2}=$16.1 K \cite{Riberolles2021,Soh2023,Donoway2024}. According to the neutron diffraction results \cite{Riberolles2021}, in both magnetic phases the Eu magnetic moments are confined to the hexagonal \textit{ab}-plane.
Below $T_{N1}$, they form an amplitude modulated magnetic phase (AMMP) with the propagation vector $Q_1$ \cite{Riberolles2021,Soh2023,Masaki2025}. 

In this state, the magnitude of magnetic moments varies sinusoidally along the $c$-axis.
In turn, below $T_{N2}$, the structure transforms into a broken-helical (B-helix) one, defined by the propagation vector $Q_2$. In the B-helix phase, the directions of Eu spins in the adjacent \textit{ab}-planes differ by an angle within the range $60^\circ<\delta\phi<180^\circ$ \cite{Riberolles2021,Soh2023}. In strong magnetic fields applied perpendicular to the helical axis (\textit{c}-axis of the crystal), the compound adopts a fan-type magnetic structure \cite{Soh2023,Takeda2024}. Basically, the magnetic structure is always co-planar with spins lying in the \textit{ab}-plane if there is no component of the magnetic field along the \textit{c}-axis.
The trigonal crystal structure of EuIn$_2$As$_2$ implies the formation of three types of antiferromagnetic domains related by $C_3$ symmetry. Below $T_{N2}$, each domain has the B-helix structure in weak magnetic fields and the fan-type structure in strong fields. The structure of the domain wall (DW) is not yet known, but it is understood that it may also exhibit a similarly complex spin arrangement, which could strongly affect the transport properties determined mainly by the bulk magnetic structure, as discussed, e.g., for TbB$_4$ \cite{PhysRevB.93.174408} and EuCuSb \cite{Takahashi2020}. 
The recent resonant elastic X-ray scattering (REXS) studies corroborated the B-helix AFM transition that occurs at $T_{N2}$, but no incommensurate propagation vector was found in the lower-$T$ phase \cite{Soh2023}. 
The electrical transport measurements indicated the metallic-like nature of EuIn$_2$As$_2$  with a small concentration of hole-type carriers \cite{Yan2022}. This implies that the chemical potential in EuIn$_2$As$_2$ crosses the valence band, which contribute to the electrical transport. In such a case, the axion insulating phase may not manifest itself through a quantized Hall resistivity, in line with the experimental findings for EuIn$_2$As$_2$ \cite{Chang2023}. 

In addition, recent theoretical studies demonstrated that the A-type AFM structure of EuIn$_2$As$_2$ is altermagnetic \cite{PhysRevB.108.075150}, and the A-type AFM structure is part of the magnetic order observed in this compound. Including the complex magnetism will modify some band structure features, but the g-wave altermagnetic spin-slitting will persist \cite{pari2024nonrelativisticlinearedelsteineffect}. In the non-stoichiometric compound (hole-doped), the main electronic bands from the altermagnetic state are likely to play an important role in the electrical transport of EuIn$_2$As$_2$ \cite{PhysRevB.108.075150}.

In this paper, we report the results of our comprehensive magnetization and magnetotransport measurements performed on high-quality single crystals of EuIn$_2$As$_2$. The primary aim of this study is to understand its complex magnetic behavior, combined with the theoretical calculations. The obtained data support the B-helix AFM ground state scenario, yet also revealed a previously unnoticed magnetic field-induced transition below $T_{N2}$. The novel feature manifests itself as a distinct hysteresis in the transverse magnetoresistance isotherms and can be attributed to the transformation from low-resistive to high-resistive DW states that occurs due to the reduction of the Fermi surface sheets' intersection area in the DWs, as indicated by our electronic band calculation in EuIn$_2$As$_2$.

\vspace{12pt}
\begin{center}
\noindent\textbf{II. MATERIALS AND METHODS}  
\end{center}

\textit{Crystal growth:} Single crystals of EuIn$_2$As$_2$ were grown by self-flux method. The elements were put in a 1:12:4 ratio in an alumina crucible and sealed inside an evacuated quartz ampule. The ampule was heated very slowly to 1100$^\circ$C and kept for 24 hours, then cooled down to 800$^\circ$C at 2$^\circ$C/h kept at this temperature for 20 hours, followed by slow cooling down to 700$^\circ$C at a rate of 0.5$^\circ$C/h. Excess flux was removed with a centrifuge. Platelet-shaped EuIn$_2$As$_2$ single crystals with dimensions $\approx 3\times2\times0.2$ mm$^2$ were obtained. The back-scattering Laue diffraction (Laue-COS, Proto Manufacturing) was used to examine the crystal orientation and quality (see Supplementary Fig. S1 \cite{SM}).

\textit{Magnetic and transport measurements:} Magnetization measurements were performed on an oriented single crystal using a SQUID magnetometer (MPMS-XL, Quantum Design). Electrical transport measurements were performed, using PPMS, on a rectangular-shaped sample cut from an oriented single crystal using a wire saw, with 50-$\mu$m silver wires attached to the sample using silver epoxy paste. An electric current of 3 mA was applied along the basal \textit{ab}-plane. The longitudinal resistivity and Hall resistivity were recorded in a complete loop of the magnetic field. To correct for misalignment of the contacts, the $\rho_{xx}(B)$ and $\rho_{yx}(B)$ data were symmetrized and antisymmetrized, respectively, as follows: 
$$
\rho_{yx}^\mathrm{up}({B}) = \frac{\rho_{yx}^\mathrm{up}({B}) - \rho_{yx}^\mathrm{dn}({-B})}{2}; \quad\quad  \rho_{yx}^\mathrm{dn}({B}) = \frac{\rho_{yx}^\mathrm{dn}({B}) - \rho_{yx}^\mathrm{up}({-B})}{2}; $$
$$ 
\rho_{xx}^\mathrm{up}({B}) = \frac{\rho_{xx}^\mathrm{up}({B}) + \rho_{xx}^\mathrm{dn}({-B})}{2}; \quad\quad  
\rho_{xx}^\mathrm{dn}({B}) = \frac{\rho_{xx}^\mathrm{dn}({B}) + \rho_{xx}^\mathrm{up}({-B})}{2};$$
Where $\rho_{yx}^\mathrm{up}({B})$ ($\rho_{xx}^\mathrm{up}({B})$) and $\rho_{yx}^\mathrm{dn}({B}$) ($\rho_{xx}^\mathrm{dn}({B})$) are the swept-up transverse (longitudinal) and swept-down transverse (longitudinal) resistivities, respectively.  

\textit{Density functional theory:} We performed density functional theory-based \textit{first-principles} calculations as implemented in Quantum ESPRESSO \cite{giannozzi2009quantum}. The antiferromagnetic ground state was obtained using ultra-soft pseudopotentials under generalized gradient approximation with Perdew–Burke–Ernzerhof type of exchange-correlation functional \cite{pseudos,perdew1996generalized}. 
For the system EuIn$_2$As$_2$, we used the experimental lattice constants reported in the literature \cite{PhysRevB.108.075150}.
The kinetic energy cut-off of 65 Ry and charge density cut-off of 780 Ry was used with a uniform Monkhorst-Pack grid (\textit{k}-mesh) of 12$\times$12$\times$8 \cite{monkhorst1976special}. High electronic self-consistency convergence criteria of at least 10$^{-10}$ were followed in all the calculations. Following these calculations, we performed wannierization for all the systems using the Wannier90 code \cite{w90}. We performed postprocessing steps in Wannier90 to calculate the anomalous Hall conductivity (AHC) $\sigma_{\alpha\beta}$ as an integral of the \textit{total} Berry curvature $\Omega_{\alpha\beta}$ over the entire Brillouin zone using a dense \textit{k}-mesh of 200 $\times$ 200 $\times$ 100 with adaptive mesh refinement scheme \cite{w90-ahc}. 
In the case of altermagnets, the contribution to the AHE does not come from the entire k-space since the Berry curvature could be zero along high-symmetry points or at the border of the Brillouin zone; therefore, a denser grid is necessary. To avoid numerical inaccuracy in the wannierization without SOC and magnetization, we performed the wannierization of the \textit{s}-In and \textit{p}-As in the nonmagnetic phase without SOC and later, we added SOC and magnetism on the \textit{p}-As states, combining first principles and model Hamiltonian as described in the Supplement materials. 


\vspace{12pt}
\begin{center}
\noindent\textbf{III. FIELD-INDUCED TRANSITIONS AND MAGNETIC ANISOTROPY} 
\end{center}

\begin{figure*}
\includegraphics[width=1\textwidth]{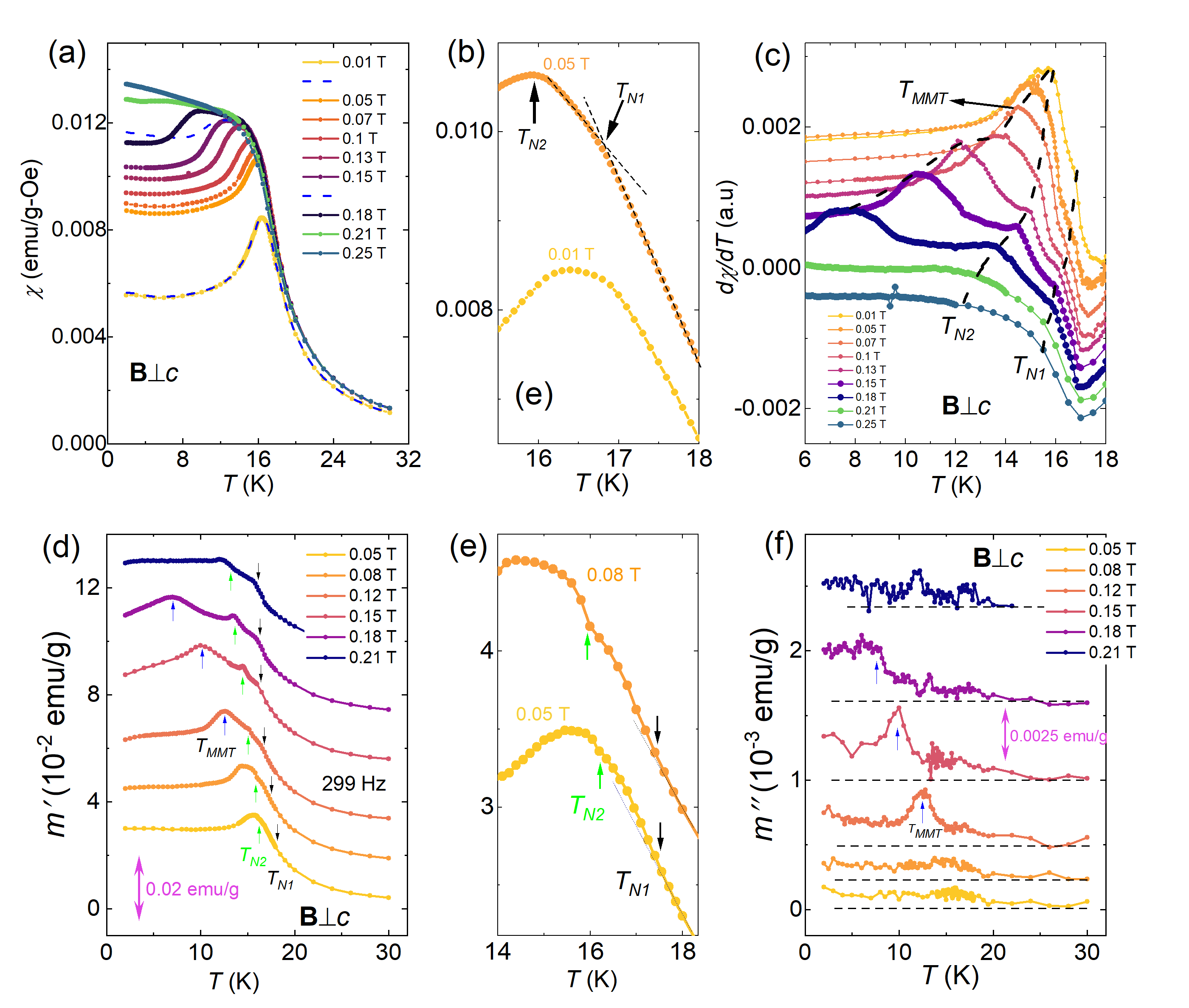}
\caption{(a) Temperature-dependent susceptibility ($\chi$) for $\mathbf{B} \perp c$, measured in the zero-field cooled under various magnetic fields. Blue dotted line represents the $\chi$ measured under field cooling for 0.01 and 0.15 T. (b) Zoom portion of (a) for the 0.01 and 0.05 T encompassing two transition temperatures ($T_{N1}$ and $T_{N2}$). (c) $d\chi/dT$ depicts the three transition temperatures: $T_{N1}$, $T_{N2}$, and $T_{MMT}$. Curves are offset for clarity. (d) The real part of the AC susceptibility measured in different DC magnetic fields, $\mathbf{B} \perp c$, with a driving field of 10 Oe and frequency of 299 Hz. (e) Zoom portion of (a) for the 0.05 and 0.08 T encompassing two transition temperatures ($T_{N1}$ and $T_{N2}$). (f) The imaginary part of the ac susceptibility measured in the same DC and AC fields. The arrows represent the transition temperatures $T_{N1}$, $T_{N2}$, and $T_{MMT}$. Data in (d) and (f) are offset by 0.02 and 0.0025 $\text{emu/g}$, respectively.}
\label{Figure1}
\end{figure*}

Figure \ref{Figure1}a presents the temperature-dependent magnetic susceptibility ($\chi$) measured under various magnetic fields for the configuration $\mathbf{B}\perp\textit{c}$. At low field (0.01 T), a broad magnetic transition is observed. With increasing field (e.g., to 0.05 T), this broad feature splits into two distinct anomalies: $T_{N1}$ = 17.6 K, identified by a change in slope (highlighted by a dotted line), and $T_{N2}$ = 16.1 K, corresponding to a maximum in $\chi$ (see Fig.~\ref{Figure1}b). The derivative of $\chi$ ($d\chi/dT$), shown in Fig.~\ref{Figure1}c, exhibits a clear kink at $T_{N1}$ and a maximum at $T_{N2}$, with the kink becoming more pronounced at intermediate fields (e.g., 0.13 T). Both transitions shift systematically toward lower temperatures with increasing magnetic field.
Additionally, a new anomaly $T_{MMT}$ emerges above 0.05 T as a broad maximum, separated from the $T_{N2}$ (see dotted line in Fig.~\ref{Figure1}c), which also shifts to lower temperatures with increasing field. Below $T_{MMT}$, a bifurcation between the zero-field-cooled (ZFC) and field-cooled (FC) $\chi$ curves becomes apparent, at 0.15 T. This bifurcation indicates the presence of a first-order metamagnetic (MM) transition.

To further investigate the magnetic transitions, temperature-dependent AC susceptibility measurements were performed. The real part of the AC magnetization ($m'$), shown in Figure~\ref{Figure1}d, confirms the presence of transitions at $T_{N1}$, $T_{N2}$, and $T_{MMT}$, all of which shift to lower temperatures with increasing magnetic field. A magnified view of the transition region is presented in Fig~\ref{Figure1}e, where $T_{N1}$ is associated with a change in slope and $T_{N2}$ appears as a subtle kink, both of which sharpen with increasing field. The $T_{MMT}$ transition begins to emerge as a maximum in $m'$ at ~0.12 T, consistent with the DC susceptibility results.
The imaginary component of the AC susceptibility ($m''$), shown in Fig.~\ref{Figure1}f, further corroborates the presence of the $T_{MMT}$ transition. This feature arises above 0.08 T, reaches a maximum, and is fully suppressed above 0.21 T. The presence of $m''$ at $T_{MMT}$ strongly supports the first-order nature of the MM transition, likely associated with a spin-flop mechanism, in agreement with the DC magnetization data.
Furthermore, field-dependent AC magnetization measurements reveal the presence of an MM state between $B^\mathrm{up}_1$ and $B^\mathrm{up}_2$ (see Supplementary Fig. S2c \cite{SM}), consistent with the DC magnetization results. A significant imaginary component in the AC magnetization is also detected within the MM region at 2 K (see Fig. S2d), coinciding with the field range in which $T_{MMT}$ is observed in the temperature-dependent $m''$ measurements.


\begin{figure}
\includegraphics[width=\columnwidth]{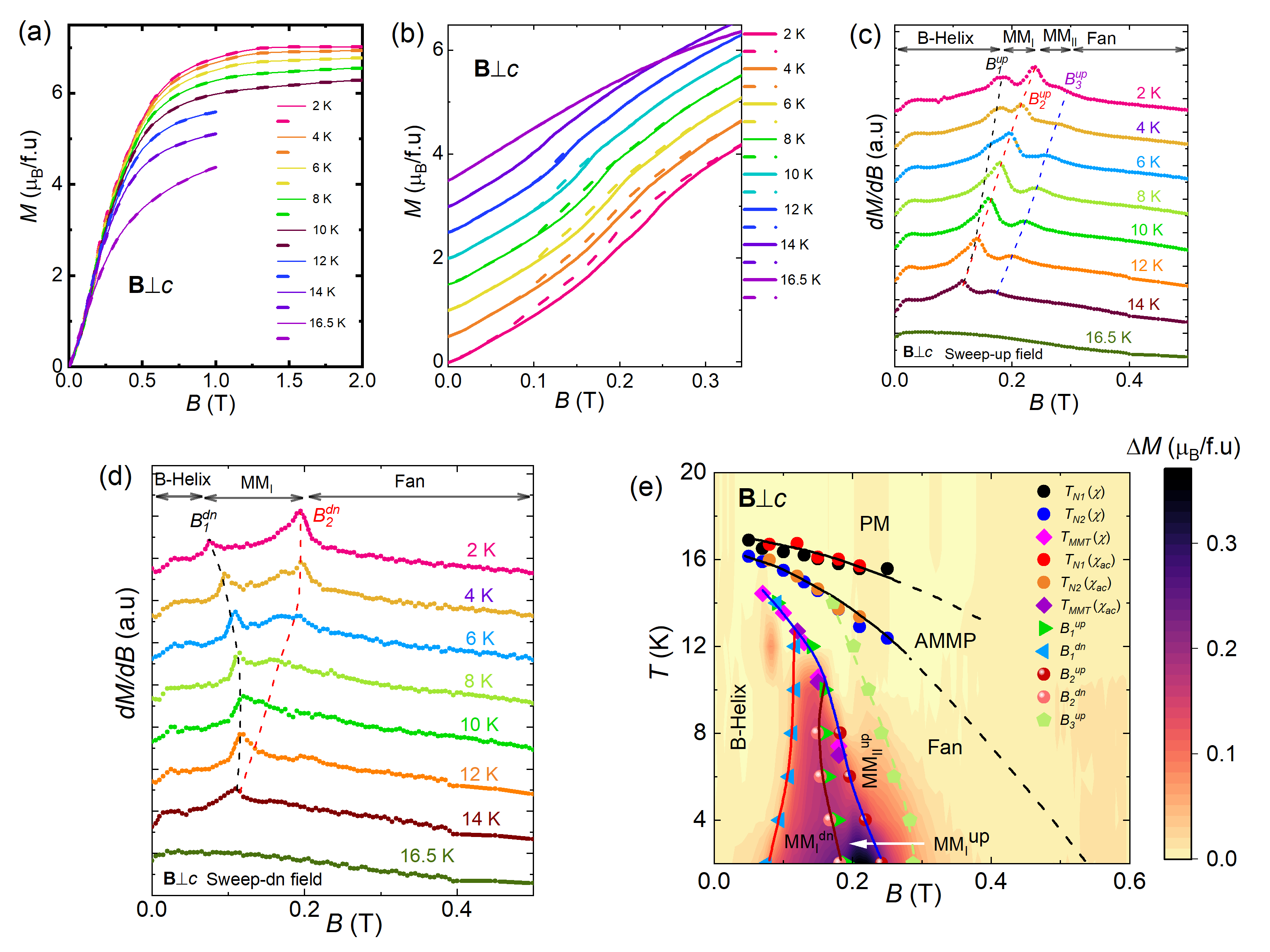}
\caption{\textbf(a) Isothermal magnetization measured at several different temperatures in swept-up field (solid line) and swept-down-field (dotted line) regimes. (b) Similar isothermal magnetization in field range 0 - 0.35T. Data at different temperatures in this plot are offset by 0.5$\mu_B$ for clarity. Derivative of magnetization (taken from Fig.~\ref{Figure2}a) plotted at different temperatures in swept-up-field (c) and swept-down-field (d). The black, red and blue dashed lines represent the $B^\mathrm{up}_1$ (or $B^\mathrm{dn}_1$), $B^\mathrm{up}_2$ (or $B^\mathrm{dn}_2$) and $B^\mathrm{up}_3$ transitions, respectively. (e) The color map of the $\Delta(M) (\equiv M^\mathrm{dn}-M^\mathrm{up})$, where 'dn' and 'up' mark the data collected in the swept-down- and swept-up-field regime, respectively. $T_{N2}$, $T_{N1}$ and $T_{MMT}$ are the transition temperatures extracted from the DC and AC magnetization data, shown in Figs.~\ref{Figure1}(b-f). $B^\mathrm{up}_1$, $B^\mathrm{up}_2$, $B^\mathrm{up}_3$ and $B^\mathrm{dn}_1$, $B^\mathrm{dn}_2$ are transition fields for swept-up and swept-down fields, respectively, extracted from Fig. \ref{Figure2}(c-d).}  
\label{Figure2}
\end{figure}

To get a deeper insight into the nature of the magnetic states, the magnetization ($M$) isotherms were measured in increasing (solid lines) and decreasing (dashed lines) magnetic field applied perpendicular to the \textit{c}-axis, $\mathbf{B}\perp\textit{c}$ (Fig.~\ref{Figure2}a). 
At 2 K, $M(B)$ is nonlinear even in weak magnetic fields and nearly saturates near $B_\mathrm{sat}$ = 0.9 T (see the Figs.~\ref{Figure2}a and ~\ref{Figure2}b). As can be inferred from Fig.~\ref{Figure2}c, the derivative d$M$/d$B$ of the data taken in the swept-up-field regime, exhibits two peaks at $B^\mathrm{up}_1$ = 0.2 T and $B^\mathrm{up}_2$ = 0.23 T, while that of data collected in the swept-down-field regime shows corresponding peaks at $B^\mathrm{dn}_1$ = 0.09 T and $B^\mathrm{dn}_2$ = 0.2 T (see Fig.~\ref{Figure2}d). In addition, we also observed an additional peak at $B^\mathrm{up}_3$ = 0.29 T for the swept-up-field.
With increasing temperature, the peak of d$M$/d$B$ at $B^\mathrm{up}_3$ shifts to lower fields, while the peaks  at $B^\mathrm{up}_2$ (or $B^\mathrm{dn}_2$) come closer to these at  $B^\mathrm{up}_1$ (or $B^\mathrm{dn}_1$), and above 8 K their positions overlap. In addition, their magnitude rapidly diminishes. Also the values of $B^\mathrm{up}_1$ and $B^\mathrm{dn}_1$ come gradually closer to each other and position near $B$ = 0.1 T at $T$ = 14 K. The evolution of the peaks in d$M$/d$B$($T$) illustrates the behavior of the hysteresis in $M$($B$) that at 
$T$ = 2 K develops near 0.1 T and vanishes in about 0.3 T, but shrinks with rising temperature and disappears near 16 K (cf. the Fig.~\ref{Figure2}b). 

The magnetization data collected for EuIn$_2$As$_2$ with $\mathbf{B}\perp\textit{c}$ are summarized in Fig.~\ref{Figure2}e in the form of a $B-T$ phase diagram. Besides the phase boundaries corresponding to the transitions from the paramagnetic state to the AMMP and then into the B-helix state, there are additional lines in this diagram which were defined by the values of $B^\mathrm{up}_1$, $B^\mathrm{dn}_1$, $B^\mathrm{up}_2$, $B^\mathrm{dn}_2$, and $B^\mathrm{up}_3$. These new phase boundaries trace an intermediate metamagnetic (MM) phases (termed (MM$^\mathrm{up}_\text{I}$, MM$^\mathrm{up}_\text{II}$) and MM$^\mathrm{dn}_\text{I}$ if observed in the swept-up-field and the swept-down-field regime, respectively). Furthermore, there is the fan-type structure state, which emerges in magnetic fields stronger than those which delineate the MM$^\mathrm{up}$ phase (see Fig.~\ref{Figure2}e). The presence of B-helix and metamagnetic transitions is consistent with previous reports \cite{Masaki2025, Alex}, which employed resonant elastic X-ray scattering (REXS) and optical polarimetry, respectively. However, in our measurements, we identify an additional metamagnetic transition (MM$^\mathrm{up}_\text{II}$) in close proximity to the previously reported one.

To understand the evolution of magnetic structures in EuIn$_2$As$_2$ with increasing magnetic field strength, it is important to realize that in the B-helix structure state, observed at low temperatures in weak magnetic fields, there are three types of AFM domains. Following Ref. \cite{Soh2023}, it can be assumed that in the field $B^\mathrm{up}_1$ the Eu spins rotate perpendicular to the magnetic field in two domains, but remain unchanged in the third one. Then, in the field $B^\mathrm{up}_2$, the Eu-spins in the latter domain flip parallel to the magnetic field, while those in the other two domains gradually turn toward the magnetic field direction. However, in the swept-up-field, an additional transition at $B^\mathrm{up}_3$ suggests that the previously proposed picture of magnetic domain wall evolution under an in-plane magnetic field may need to be amended. Full polarization of the moments in all three domains is achieved above the field $B_\mathrm{sat}$. The transition from the helical AFM to the MM state can be first- or second-order, depending on the turn angle between Eu-moments in adjacent layers \cite{Johnston2017}. The hysteretic behavior of this transition in EuIn$_2$As$_2$, accompanied by a large imaginary component in the temperature- and magnetic-field-dependent AC magnetic susceptibility of the compound in the MM state (see Fig.~\ref{Figure1}f and Supplementary Fig. S2d \cite{SM}), hints at the first-order MM transition. 

\vspace{12pt}
\begin{center}
\noindent\textbf{IV. HYSTERETIC MAGNETORESISTANCE} 
\end{center}

\begin{figure}
\includegraphics[width=1.0\columnwidth]{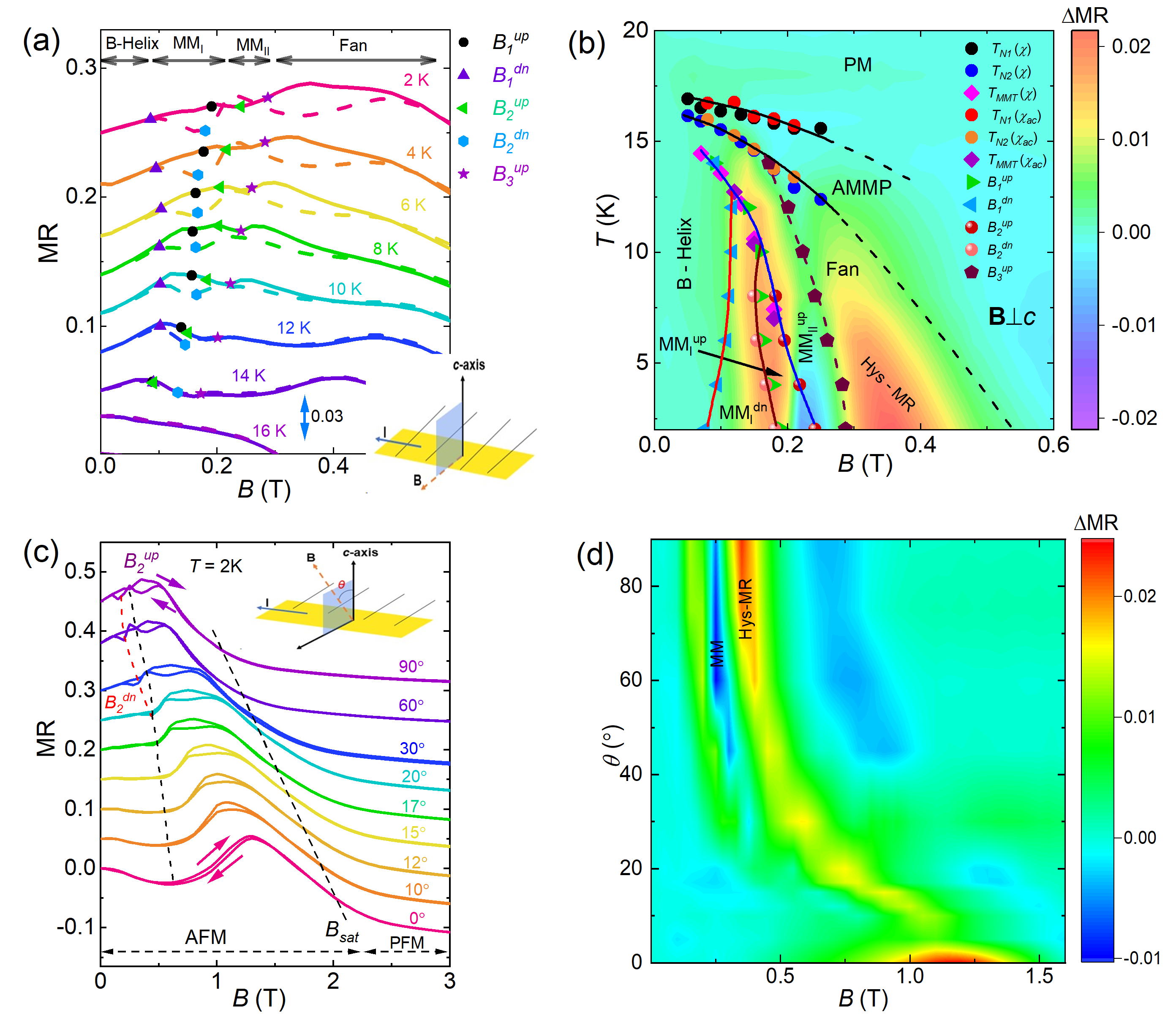}
\caption{(a) Magnetoresistance (MR) measured at different temperatures in swept-up- (solid line) and swept-down-field (dotted line). Data in this plot are offset for clarity. $B^\mathrm{up}_1$, $B^\mathrm{up}_2$, $B^\mathrm{up}_3$ and $B^\mathrm{dn}_1$, $B^\mathrm{dn}_2$ are the same characteristic fields, defined in Fig.~\ref{Figure2}(c-d). (b) The color map of the $\Delta$MR (= MR$^\mathrm{up}$-MR$^\mathrm{dn}$). (c) MR measured at different angles in swept-up and swept-down. Data are offset by 0.05 for clarity. $B^\mathrm{up}_2$ and $B^\mathrm{dn}_2$ are transition fields for swept-up and swept-down fields, respectively. $B_\mathrm{sat}$ are saturation fields interpolated between values determined for $\textbf{B}\parallel c$  and for $\textbf{B}\perp c$. The inset shows the diagram of MR measurement. (d) The color map of the $\Delta$MR$(B,\theta)$, showing the hysteresis in the MM and fan-type structure states.} 
\label{Figure3}
\end{figure}

We determined the magnetoresistance (MR=$[\rho_{xx}(B)/\rho_{xx}(0)]-1$) for $\mathbf{B}\perp\textit{c}$, at various temperatures. Results are shown in Fig.~\ref{Figure3}a. Measurements conducted in swept-up and swept-down fields lead to slightly different results, showing the hysteresis of MR in the field range where the compound undergoes magnetic field-induced phase transitions. 
Notably, at temperatures below 10\,K MR measured for the swept-up-field is clearly larger compared to the swept-down-field MR in two ranges of field:  around 0.15 T and around 0.35 T. Fig.~\ref{Figure3}b depicts the $B-T$ phase diagram superimposed on a color map of $\Delta$MR (= MR$^\mathrm{up}$-MR$^\mathrm{dn}$), indicating the presence of hysteretic MR in both MM and fan-type structure states.  
Nevertheless, a key observation is that at low temperatures and for $B>B^\mathrm{up}_3$ (i.e. in the fan-type state) the MR exhibits significant hysteresis, even though this is not reflected in the magnetization (see Figs.~\ref{Figure2}e and \ref{Figure3}b for comparison). 
The hysteretic MR (Hys-MR) could arise from the complex magnetic structures in the DWs \cite{PhysRevB.93.174408} or from the competitive nature of the ground state at the boundary of the collinear phase and helical phase within the domains, like in EuCuSb and EuZnGe       
 \cite{Takahashi2020, Kurumaji2022}. 
The latter conjecture is ruled out for EuIn$_2$As$_2$, as its ground state is the homogeneous B-helical phase, without phase boundaries.  
Thus, we argue that Hys-MR in EuIn$_2$As$_2$ emerges due to DWs' contribution to the scattering of charge carriers. Furthermore, DWs can be tuned between high and low resistivity, depending on the carrier concentration, and the strength of interaction of conduction electrons with the local magnetic moments \cite{Suzuki2019} or the spin configuration in the magnetic structure \cite{Ueda2015, Ueda2014}. 

Next, we investigated the MR at 2 K, for different angles $\theta$ between \textit{c}-axis and magnetic field, rotating it from $\mathbf{B}\perp\textit{c}$ to $\mathbf{B}\parallel\textit{c}$,  as shown in Fig.~\ref{Figure3}c (the current was always perpendicular to the magnetic field). By rotating the magnetic field towards \textit{c}-axis, the slope of low-field positive MR decreases and reaches zero at around  $\theta=$15$^{\circ}$. For $\theta<$15$^{\circ}$ negative MR arises and becomes strongest at  $\theta=$0$^{\circ}$. 
The negative MR is due to the gradual alignment of the Eu moments with increasing magnetic field, reducing the spin-disorder electron scattering. However, above a specific field, $B^\mathrm{up}_2$, MR upturns and reaches a maximum, marking the occurrence of an additional scattering mechanism. 
In even higher fields, this additional contribution to the scattering gradually fades out and becomes negligible in fields close to $B_\mathrm{sat}$. 

The color map of the $\Delta$MR$(B,\theta)$ in Fig.~\ref{Figure3}d, shows that in the MM state, hysteresis is suppressed for $\theta < 30^{\circ}$, but the Hys-MR regime persists even below the 30$^{\circ}$, indicating that both have a different source of electron scattering. 
The hysteresis within the MM state arises due to the first-order nature of the MM transition, which is consistent with the isothermal magnetization. Indeed, above the $B^\mathrm{up}_2$, the Hys-MR, along with increasing MR, can have numerous mechanisms, for example, proposed in literature, such as Fermi-surface reconstruction \cite{Marcus2018} or electronic wave localization \cite{Seo2021}. 
In our bulk magnetization measurements, $B^\mathrm{up}_2$ transition is not observed for $\mathbf{B}\parallel\textit{c}$ (see Supplementary Fig. S2b \cite{SM}), but it manifests in the MR as a minimum (see Fig.~\ref{Figure3}c). 
It implies that the transition at $B^\mathrm{up}_2$ is not caused by the bulk magnetic structure. 
In addition, our electronic band structure calculations, combined with the electrical transport measurements, show that the material is a low-carrier-concentration metal with an easily tunable electronic structure near the Fermi level (discussed in detail below). Thus, the Hys-MR, along with increasing MR, can be associated with the electron scattering on DWs, which is tunable between the low resistivity and high resistivity depending on the geometry of the electronic or magnetic structure \cite{Suzuki2019, Ueda2014, Ueda2015}.  
For example, in pyrochlores, the antiferromagnetic DWs were reported to be metallic for all-in-all-out spin configurations and insulating for two-in-two-out spin configurations \cite{Ueda2015, Ueda2014}. But this form of spin-configuration does not exist in the EuIn$_2$As$_2$. 

 

\vspace{12pt}
\begin{center}
\noindent\textbf{V. ANOMALOUS HALL EFFECT}
\end{center}

To further understand the origin of the observed hysteresis in MR and its interplay with the electronic structure, we examined the Hall resistivity ($\rho_{yx}$) shown in the Figs.~\ref{Figure4}a and b. These measurements were recorded at different angles of applied magnetic field, rotated from $\mathbf{B}\parallel\textit{c}$ to $\mathbf{B}\perp\textit{c}$, at 2 K and 100 K, respectively, both in swept-up and swept-down field regimes, showing no hysteresis. For $\theta<90^\circ$ and fields above $B_\mathrm{sat}$, we fitted $\rho_{yx}$ with the following equation to estimate the carrier concentrations:  
\begin{equation}
\rho_{yx} = R_0B\cos\theta, 
\label{eqnS2}
\end{equation}
where $R_0$ = 1/${n_e}e$, $\theta$ is the angle between $\mathbf{B}$ and \textit{c}-axis. Inset of Fig.~\ref{Figure4}b shows the angle-dependent carrier concentrations, which are hole-type. 

It is noted that carrier concentration is nearly the same at 2 K and 100 K (see to inset Fig.~\ref{Figure4}b and Supplementary Fig. S8b \cite{SM}), therefore we obtained $\Delta\rho_{yx}$ by subtracting the ordinary Hall resistivity from the total Hall resistivity, using the relation: $\Delta\rho_{yx}(2K)$ = $\rho_{yx}(2K)-\rho_{yx}(100 K)$, 
where $\rho_{yx}$(2 K) and $\rho_{yx}$(100 K) are the Hall resistivities measured at temperatures 2 K and 100 K, respectively (see Figs.~\ref{Figure4}a and the \ref{Figure4}b), and {$\Delta\rho_{yx}$ is shown in the Fig.~\ref{Figure5}a. As reported in \cite{Yan2022}, the maxima in $\Delta\rho_{yx}$ within the B-helix state are attributed to the non-collinear magnetic structure in this compound.

\begin{figure*}
\includegraphics[width=0.8\textwidth]{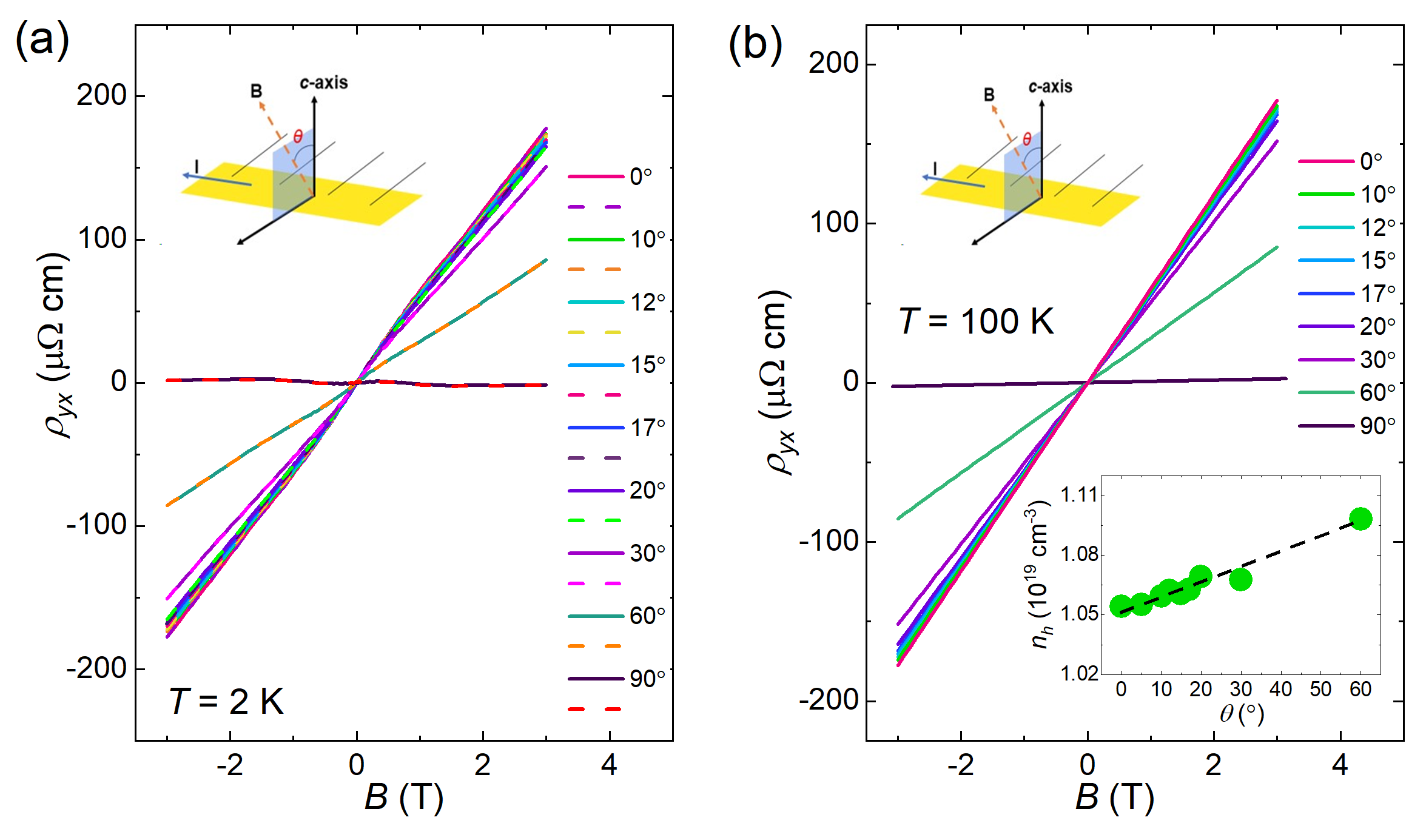}
\caption{Hall resistivity as a function of magnetic field recorded at various angles for (a) $T$ = 2 K and (b) $T$ = 100 K. Insets to (b) display the angle-dependent carrier concentration at 100 K, respectively. The carrier concentration was estimated using equation \ref{eqnS2}.}
\label{Figure4}
\end{figure*}
 
\begin{figure}
\includegraphics[width=\columnwidth]{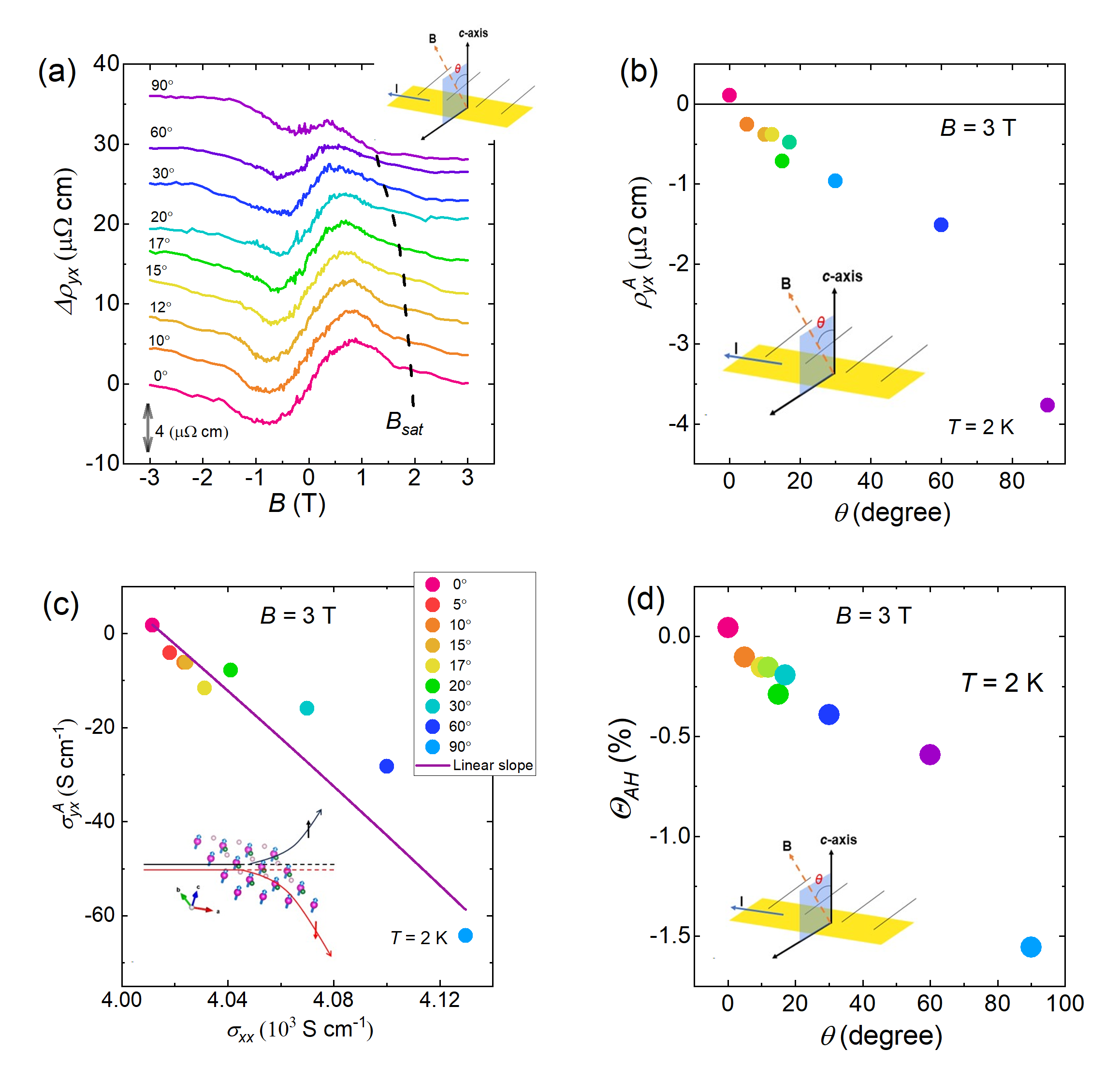}
\caption{(a) The residual Hall resistivity, $\Delta\rho_{yx}$, at 2 K, plotted for different field angles. Data are offset for clarity. (b) Angular dependence of $\rho^A_{yx}$ in a field of 3 T, i.e., above the saturation field. (c) Anomalous Hall conductivity $\sigma^A_{yx}$ plotted against the longitudinal conductivity $\sigma_{xx}$ in a field of 3 T for various angles. The straight violet line is a guide for the eyes, indicating that $\sigma^A_{yx}$ follows the linear relationship with $\sigma_{xx}$. Inset: schematic representation of asymmetric scattering of conduction electrons on the local moments of Eu. (d) The anomalous Hall angle, $\Theta_{AH}$, vs. field orientation angle, $\theta$, in a field of 3 T. Inset shows the diagram of the Hall measurement.}
\label{Figure5}
\end{figure}


\begin{figure}[b!] 
    \includegraphics[height=0.4\textheight]{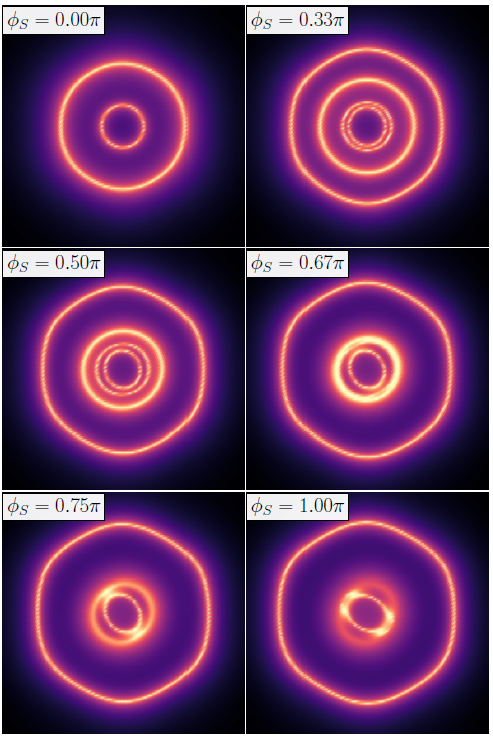}
    \caption{Evolution of the Fermi Surface around the $\Gamma$ point as spin alignment changes from A-type AFM ($\phi_S=0$) to FM with spin in the plane ($\phi_S=\frac{\pi}{2}$) for the energy $\omega$=-150 meV below the Fermi level expected for a stoichiometric compound.}
    \label{Fig4:FS_inplanefield}
\end{figure}

Above $B_\mathrm{sat}$, the system enters a fully polarized state and $\Delta\rho_{yx}$ corresponds to the anomalous Hall resistivity ($\rho^A_{yx}$). Fig.~\ref{Figure5}b shows the angle dependence of $\rho^A_{yx}$. At 0$^\circ$ and 2 K, $\rho^A_{yx}$ is small. As the angle increases, -$\rho^A_{yx}$($T$) grows, reaching 3.7~$\mu\Omega$cm at 90$^\circ$.

In polarized ferromagnetic systems, \KS{$\rho^A_{yx}$} can be described as \cite{Nagaosa2010}:
\begin{equation}
\rho^A_{yx} = R_s \mathrm{M}_z
\label{eq1}
\end{equation}
where $R_s$ represents the anomalous Hall coefficient coupled to M$_z$, which can be intrinsic or extrinsic. For intrinsic and extrinsic transverse scattering, $R_s$ is proportional to the $\rho^2_{xx}$ and $\rho_{xx}$, respectively \cite{Nagaosa2010}. 


Next, we convert it to anomalous Hall conductivity (AHC) by following the relation: $\sigma^A_{yx}=\rho^A_{yx}/\rho^2_{xx}$, since $\rho^A_{yx}\ll\rho_{xx}$. Above $B_\mathrm{sat}$, the local $4f$ moments are aligned with the field, and remain the same for all angles between $\mathbf{B}\parallel\textit{c}$ and $\mathbf{B}\perp\textit{c}$, but it should be noted that $\sigma^A_{yx}$ does not remain constant (see Fig.~\ref{Figure5}c). It is because the coefficient $R_s$ depends on $\rho_{xx}$, causing the $\sigma^A_{yx}$ to vary with the  $\sigma_{xx}(=\rho_{xx}/\rho^2_{xx})$. For $\theta=$0$^\circ$, $\sigma^A_{yx}$ is 
equal to the 1.83~Scm$^{-1}$. With increasing the angle, $\sigma^A_{yx}$ increases,
and reaches -63~Scm$^{-1}$ for 90$^\circ$. The $\sigma^A_{yx}$ follows the $\sigma_{xx}$ in linear manner (cf. Fig.~\ref{Figure5}c). This underpins the argument of itinerant electrons' asymmetric scattering induced by the polarized local $4f$ moments (sketched in the inset of Fig.~\ref{Figure5}c) \cite{Nagaosa2010}.
Finally, we calculate the Hall angle ($\Theta_{AH}$) defined as: 

\begin{equation}
\Theta_{AH} = \sigma^A_{xy}/\sigma_{xx}.
\label{ThetaAH}
\end{equation}

Fig.~\ref{Figure5}d shows the Hall angle for different $\theta$, reaching the value of -1.6$\%$ at 90$^\circ$, which is small, as expected for topologically trivial electronic structure with low spin-orbit coupling \cite{Nagaosa2010}.


\vspace{12pt}
\begin{center}
\noindent\textbf{VI. ELECTRONIC STRUCTURE CALCULATIONS}  
\end{center}
\hspace{2cm}

To shed more light on the electronic structure, we performed density functional theory (DFT) calculations starting from the collinear antiferromagnetic state with propagation vector $\tau=$ (0,~0,~1) and rotating the spin under the effect of the external magnetic field. Therefore, we computed the evolution of the Fermi surface under the in-plane magnetic field by tuning the azimuthal angle of the spins $\phi_S$, while to simulate the out-of-plane magnetic field, we tuned the out-of-plane magnetic field (more details are reported in Supplement Materials \cite{SM}). In our notation, $\phi_S$=0 corresponds to the A-type AFM with in-plane spins, while  $\phi_S$=$\pi$ corresponds to the FM phase with in-plane spins.

We study the Fermi surface for k$_z$=0 at the energy $\omega$=-150 meV to simulate the p-type Fermi level, which is reported in Fig. \ref{Fig4:FS_inplanefield}. We obtain 4 Fermi sheets at zero magnetic field corresponding to $\phi_S$=0, we have two degenerate Fermi surfaces since we are in the nodal plane of the altermagnet. A value of $\phi_S$ different from zero breaks Kramer's degeneracy, pushing the system to a nodeless band structure \cite{PhysRevB.109.024404} and we observe 4 Fermi surfaces for $\phi_S$=0.33$\pi$ and 0.50$\pi$. Increasing to $\phi_S$=0.67$\pi$, there is a Lifshitz transition for which 1 FS gaps out at the $\Gamma$ point and we are left with 3 FS sheets.
Increasing the in-plane magnetic field corresponds to increasing $\phi_S$, until the system goes towards a fully ferromagnetic phase at $\phi_S$=$\pi$.  We repeat the same calculation at the same energy $\omega$=-150 meV for an out-of-plane magnetic field and we do not observe any Lifshitz transition (see Fig. S13 \cite{SM}).
Therefore, we have a Lifshitz transition that depends on the magnetic field orientation. For the same orientation for which we theoretically observe the Lifshitz transition, we observe the metamagnetic transition in the experimental results. 


We examined the magnetic contribution from the As-\textit{p}-orbital (at $4f$ Wyckoff position) and found that the moment of \textit{p}-orbital is antiparallel to the neighboring Eu moments (see Supplementary Figs. S9a and S9b \cite{SM}). The Fermi level (approximated from the experimental ordinary Hall resistivity) is located in the valence band. A-type antiferromagnetic coupled $4f$ moments of Eu$^{2+}$ exhibit  the altermagnetic spin-splitting feature in momentum-space  
\cite{PhysRevB.108.075150}. 


For $\theta_S=\pi$/2 (the angle between the As-spin moment and the \textit{c}-axis), the structure of the As-spins is an A-type AFM. The electronic band structure shows non-relativistic spin-splitting, while the electronic bands don't cross the Fermi level at the $\Gamma$ point. When the spin's rotation angle $\theta_S$ decreases, there is a spin-splitting at the $\Gamma$ point and $S_Z$ grows. 
In addition, the results of our calculations show that in an applied magnetic field, the spin-splitting from the magnetic field also occurs at the $\Gamma$-point (see Fig. S11f \cite{SM}) and could contribute to magnetotransport. In the presence of As-spin-canting that could be due to the magnetic field via the interaction with the local $4f$ moments or relativistic weak ferromagnetism, the size of the spin-splitting increases and the bands cross the Fermi level (see Fig. S11 \cite{SM}). 
Following the theoretically calculated electronic structure, the longitudinal resistivity could be reduced. Nevertheless, we have observed an upturn of the MR  in $B^\mathrm{up}_2$, which suggests that growing $S_Z$ of As-spin via interaction with the local 4f moments could reduce the intersection area of Fermi surface sheets ($A_\mathrm{int}$) on the DWs and increase the electron scattering. 
  
Next, the AHC is calculated for the respective rotation angles (see Fig. S11a-f) \cite{SM}, lower, right-side panel). Due to the topological band structure in the ferromagnetic phase, we have opposite values of the AHE in the valence and conduction band due to the opposite values of the Berry phase in topological materials \cite{PhysRevB.98.125408}. For $\theta_S$ = 0, a finite AHC of the order of 3 Scm$^{-1}$ is obtained at the Fermi-level, which is quite close to the experimental value of 1.8 Scm$^{-1}$ for $\theta$ = 0 (see Fig.~\ref{Figure5}c). Thus, it shows that As-spin contributes to the AHC with a small positive value, and supports that it plays the main role in DWs scattering. We interpreted the AHC in the polarized FM state of the local $4f$ moment, where the theoretical scenario $\theta_s$ = 0 applies for B$>B_{sat}$. Due to the relatively simple structure of the AHE, which is positive in the conduction band and negative in the valence band and to the impossibility of having strong band structure reconstruction, we can assume that the intrinsic AHC stays small and positive for the different orientations of the magnetic field.

\vspace{12pt}
\begin{center}
\noindent\textbf{VII. DISCUSSION}  
\end{center}
\hspace{2cm}

EuIn$_2$As$_2$ shows two antiferromagnetic transitions at $T_{N1}$ and $T_{N2}$, which correspond to the AMMP and B-helix states. With the magnetic field applied in the $ab$-plane, the magnetic state changes first to a metamagnetic state, then to a fan-like structure, and finally to a fully polarized ferromagnetic state. Notably,  the metamagnetic transition occurs at a lower field when the field is swept-down compared to when it is swept-up, pointing to a dynamic nature of the magnetic structure. MR data are consistent with the magnetization results, but they also exhibit additional features. In particular, the fan state shows a strong hysteresis in MR that is not occurred in the isothermal magnetization measurement. Angle-dependent MR further shows that the metamagnetic transition weakens as the field direction is tilted toward the $c$-axis, while the hysteresis in the fan-state persists. Moreover, an upturn in MR is observed for field angles below about 30$^\circ$. The combination of strong hysteresis and enhanced scattering in the MR cannot be fully explained by the bulk magnetic structure. Instead, these features likely arise from domain-wall dynamics.

Domain walls can significantly influence charge transport in materials with complex magnetic and electronic structures. According to the theory given in the literature \cite{Suzuki2019}, the contribution of scattering on DWs to the resistivity is inversely proportional to $A_\mathrm{int}$, where $A_\mathrm{int}$ is the intersection area shared by the Fermi surface sheets on the DWs. The Landauer-B{\"u}ttiker approach indicates that electrons can only transmit over the DWs if their transverse momentum is preserved at the Fermi level, which requires that the sheets of the Fermi surface from two neighboring domains have the intersection of finite area ($A_\mathrm{int}$) \cite{Nguyen2006, Kobayashi2018}. 
That intersection can be controlled by carrier concentration, via changing the size of the Fermi surface sheets on either side of DWs. The lower carrier concentration indicates the presence of a smaller sheet of the Fermi surface, which results in a decrease of $A_\mathrm{int}$ in the DWs and in increased electron scattering. So far, domain wall scattering phenomena have only been explored for Weyl or nodal semimetals because they have the nodes at the Fermi level associated with the low carrier concentrations of the order of 10$^{20}$cm$^{-3}$ \cite{Kobayashi2018, Suzuki2019}. Here, we expanded this phenomenon to the non-collinear magnet EuIn$_2$As$_2$, with a carrier concentration of the order of 10$^{19}$cm$^{-3}$,  about 10-fold lower than in magnetic Weyl semimetal CeAlGe \cite{Suzuki2019}. In EuIn$_2$As$_2$, the local 4\textit{f}-level is distant from the Fermi level. Theoretical calculations show that the \textit{p}-orbital of the As atom contributes to states at the Fermi level and that electronic bands are spin-split due to altermagnetism and the induced moment on \textit{p}-orbital by Eu-4\textit{f}-moments (see Fig. S11 \cite{SM}). As a result, this spin-splitting modifies the size and geometry of the sheets of the Fermi surface (see Supplementary Fig. S13 \cite{SM}). Thus, it could also reduce the intersection area ($A_\mathrm{int}$) in the DWs around $B^\mathrm{up}_2$ (or $B^\mathrm{dn}_2$), and consequently, increase the  carrier scattering.

Theoretical calculated AHC for the electronic band structure of the As $p$-orbital yields a small finite value, which agree with the experimental AHC for $\theta$ = 0. This supports the interpretation that the underlying electronic band structure is trivial,  due to the position of the Fermi level in the valence band where the relevant spin-orbit coupling originates from As-atoms \cite{PhysRevB.108.075150}. These AHC results shed light on the contribution of the $p$-orbitals to the electrical transport, and reveal that the hysteresis observed in magnetoresistance arises from domain walls coupled to the $p$-orbitals of As.

\vspace{12pt}
\begin{center}
\noindent\textbf{VIII. CONCLUSIONS}  
\end{center} 

Combining theoretical and experimental results and taking into account magnetic and topological properties, we considered the impact of complex magnetism of EuIn$_2$As$_2$ on its magnetoresistance.
We described the metamagnetic transition associated with a large hysteresis in magnetoresistance, which turns into a sharp upturn for the magnetic field tilted by 15$^\circ$ from the \textit{c}-axis of the crystal. Our findings explain the origin of strong MR scattering associated with the hysteresis that arises from DWs. The application of a magnetic field changes the low-resistivity antiferromagnetic domain walls to the high-resistivity domain walls due to the reduction in the Fermi surface sheets interaction area in the domain walls, originating from \textit{p}-orbitals of As.

\vspace{12pt}
\begin{center}
\noindent\textbf{ACKNOWLEDGMENTS}
\end{center} 

The authors acknowledge Tomasz Dietl for useful discussions.
This work was supported by the National Science Centre (NCN, Poland) under Project no. 2021/41/B/ST3/01141 (K.S., O.P., T.R., T.T., P.W. and D.K.). J. S. and C. A. were supported by the “MagTop” project (FENG.02.01-IP.05-0028/23) carried out within the “International Research Agendas” programme of the Foundation for Polish Science co-financed by the European Union under the European Funds for Smart Economy 2021-2027 (FENG). J. S. is supported by the Narodowe Centrum Nauki (NCN, National Science Centre, Poland) Project No. 2019/34/E/ST3/00404.
We acknowledge the access to the computing facilities of the Interdisciplinary Center of Modeling at the University of Warsaw, Grant g91-1418, g91-1419, g96-1808 and g96-1809 for the availability of high-performance computing resources and support. We acknowledge the CINECA award under the ISCRA initiative  IsC99-"SILENTS”, IsC105-"SILENTSG", IsB26-"SHINY" and IsCb7-"CHARPHEN" grants for the availability of high-performance computing resources and support. We acknowledge the access to the computing facilities of the Poznań Supercomputing and Networking Center Grant No. pl0223-01 and pl0267-01.

\hspace{1cm}
\bibliography{biblio.bib}
\hspace{2cm}

\clearpage
\newpage
\onecolumngrid

\begin{center}
  \textbf{\Large Supplemental Material\\
  for\\
  Nature of field-induced transitions and hysteretic magnetoresistance \\ in the noncollinear antiferromagnet EuIn$_2$As$_2$}\\[.2cm]
 
 Karan Singh$^{1}$, Jan Skolimowski$^{2}$, Giuseppe Cuono$^{3}$, Raghottam M. Sattigeri$^{4}$, A. Ptok$^{5}$, O. Pavlosiuk$^{1}$, Tetiana Romanova$^{1}$, Tomasz Toliński$^{6}$, P. Wi\'{s}niewski$^{1}$, Carmine Autieri$^{2}$ and D. Kaczorowski$^{1}$\\[.2cm]
  {\itshape
  \mbox{$^{1}$Institute of Low Temperature and Structure Research,}\\ 
  \mbox{Polish Academy of Sciences, Okólna 2, 50-422 Wrocław, Poland}\\
  \mbox{$^{2}$International Research Centre MagTop, Institute of Physics,}\\
  \mbox{Polish Academy of Sciences, 
  Aleja Lotnik\'ow 32/46, PL-02668 Warsaw, Poland}\\

  \mbox{$^{3}$Consiglio Nazionale delle Ricerche (CNR-SPIN),}\\ 
  \mbox{Unit\'a di Ricerca presso Terzi c/o Universit\'a “G. D’Annunzio”, 66100 Chieti, Italy}\\

  \mbox{$^{4}$Physics Department, Università degli Studi di Milano, Via Celoria 16,
20133 Milan, Italy}\\
  \mbox{$^{5}$Institute of Nuclear Physics, Polish Academy of Sciences,}\\
  \mbox{W. E. Radzikowskiego 152, PL-31342 Kraków, Poland}\\
  \mbox{$^{6}$Institute of Molecular Physics, Polish Academy of Sciences,}\\
  \mbox{M. Smoluchowskiego 17, 60-179 Poznań, Poland}
  }

(Dated: \today)
\\[1cm]
\end{center}

\setcounter{equation}{0}
\renewcommand{\theequation}{S\arabic{equation}}
\setcounter{figure}{0}
\renewcommand{\thefigure}{S\arabic{figure}}
\setcounter{section}{0}
\renewcommand{\thesection}{S\arabic{section}}
\setcounter{table}{0}
\renewcommand{\thetable}{S\arabic{table}}
\setcounter{page}{1}

\newpage
\begin{figure*}[h]
\includegraphics[width=0.5\textwidth]{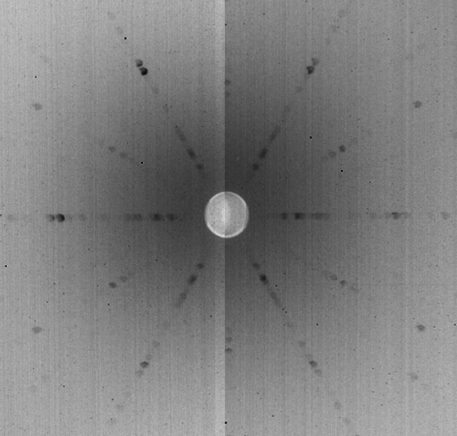}
\caption{Back-scattered Laue pattern of the EuIn$_{2}$As$_{2}$ crystal oriented along the (001) direction.}
\label{FigS1_Laue}
\end{figure*}

\begin{figure*}
\includegraphics[width=1\textwidth]{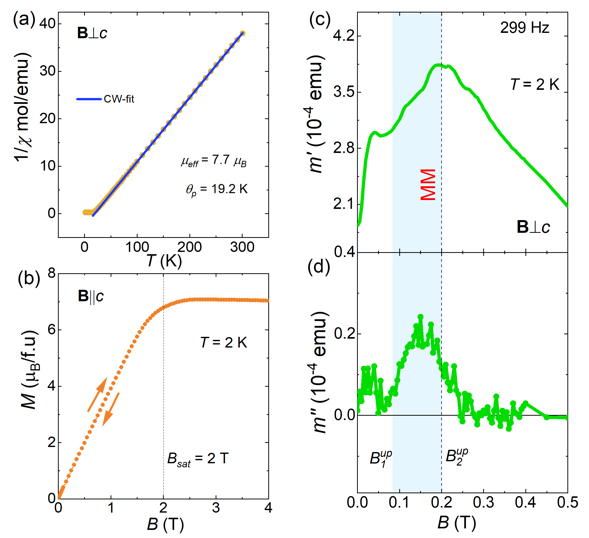}
\caption{(a) The Curie-Weiss law fit (blue line) for the $\chi^{-1}$, measured at 0.1 T, above $T_{N1}$. (b) Magnetization as a function of magnetic field recorded at 2K in a complete loop for $\mathbf{B} \parallel c$. The dotted line marks the saturation field ($B_\mathrm{sat}$). (c) and (d) The real- and imaginary-ac susceptibility as a function of the magnetic field, measured at 2 K. MM is the metamagnetic region.}
\label{magnetization}
\end{figure*}

\subsection*{1. Magnetization}

For EuIn$_2$As$_2$, inverse $\chi$ follows the Curie-Weiss law in the paramagnetic region, which was measured for $\mathbf{B}\perp\textit{c}$ at 0.1 T:

\begin{equation}
\chi = \textit{C}/(T-\theta_p)
\label{eqnS1}
\end{equation}

Where $\textit{C}$ is the Curie constant, and $\theta_p$ is the paramagnetic Curie temperature. These parameters are derived from the fitting of the equation~\ref{eqnS1} (shown as a blue line in Fig.~\ref{magnetization}a). The effective magnetic moment ($\mu_{eff}$) calculated from the parameter $\textit{C}$ is 7.7 $\mu_B$, which closely matches with the theoretical value of 7.94 $\mu_B$ for Eu$^{2+}$ ions. The obtained Curie temperature ($\theta_p$) is 19.2 K, indicating the presence of a predominance of ferromagnetic exchange interactions in the antiferromagnetic compound. Magnetization ($\textit{M}$) as a function of the magnetic field in a complete loop was measured at 2 K for $\mathbf{B}\parallel\textit{c}$ and indicates that hysteresis is absent, as illustrated in Fig.~\ref{magnetization}b. As the magnetic field increases, the magnetization increases and reaches a saturation value of 7$\mu_B$ at field $B_\mathrm{sat}$ = 2 T. This value is consistent with the theoretical moments for Eu$^{2+}$, which correspond to spin magnetic moments since the orbital angular momentum is negligible. The saturation field is larger compared to the one obtained with the field perpendicular to the $\textit{c}$-axis.

\subsection*{2. Magnetoresistance}
Fig.~\ref{FigS4} shows the temperature-dependent resistivity ($\rho_{xx}$) for $\mathbf{B}\parallel\textit{c}$, recorded at 2K and 0T. The derivative of $\rho_{xx}$ shows the two antiferromagnetically transitions $T_{N1}$ and $T_{N1}$ (see inset to Fig.~\ref{FigS4}), which corroborate the magnetization results. 
In this section, we analyze the low-field magnetoresistance, MR (= $\frac{\rho(B)-\rho(0)}{\rho(0)}$) data. Fig.~\ref{FigS5}a  shows the average MR (MR$_{av}$ = [MR$_{dn}$ +MR$_{up}$]/2) at various temperatures for $\mathbf{B}\parallel\textit{c}$. At 16 K, the MR$_{av}$ increases with increasing magnetic field and reaches a maximum at 0.5 T. Above 0.5 T, it decreases. This maximum position shifts to higher magnetic fields with decreasing temperature. The decrease in MR$_{av}$ above the maximum is attributed to the polarization of the Eu moments aligning with the applied field, resulting in reduced electron scattering. Below the maximum MR$_{av}$ at 2K, the MR$_{av}$ exhibits a minimum at 0.6 T, which shifts towards the lower field and is suppressed above 8K. In the main text, we have explained that the minimum occurs owing to domain and domain wall scattering, which is balanced at the minima field. With the rising temperature, domain wall scattering increases and becomes dominant above 8K, suppressing the minima field. 

Next, we examined magnetoresistance at various angles at 2K (Fig.~\ref{FigS5}b). Interestingly, the MR$_{av}$ exhibits a positive linear dependence on the magnetic field for angles exceeding 10$^\circ$. For 90$^\circ$, we measured the MR$_{av}$ at different temperatures (Fig.~\ref{FigS5}c), and linear MR$_{av}$ persists up to 14 K and the slope remains positive. The linear magnetoresistance at high fields (where $\omega_c\tau \gg 2\pi$, here, $\omega_c$ represents the cyclotron frequency, and $\tau$ symbolizes the relaxation time) has established explanations related to electronic properties (like lowest spin-split landau level \cite{Abrikosov1998}, low carrier mass Dirac surface state \cite{Abrikosov2000}, magnetic breakdown \cite{Naito1982} etc.) or geometrical factors (strong mobility fluctuation \cite{Song2015}, disorder \cite{Xu1997} etc.). However, its occurrence at low fields ($\omega_c\tau\ll2\pi$) is unexpected. The authors of reference \cite{Feng2019} discuss potential origins such as a change of Fermi-surface topology or scattering of charge density wave fluctuations at specific Fermi surface points. This suggests that itinerant electrons experience more abrupt scattering events on the Fermi surface before each collision arising from the change in the Fermi-surface area in sharp edges, which depends on the local magnetic moment configurations. In this scenario, a large MR$_{av}$ is expected \cite{Feng2019}. As shown in Fig.~\ref{FigS5}c, the MR$_{av}$ reaches 2$\%$ at a relatively low applied magnetic field of 0.1 T. This large value at low field supports a changeover of Fermi surface topology turning to a strong scattering. Furthermore, we measured the MR in high magnetic fields up to 9 T for $\mathbf{B} \parallel c$ (Fig.~\ref{FigS6}a) and $\mathbf{B}\perp c$ (Fig.~\ref{FigS6}b), where it reaches a larger value of 63 and 65$\%$, respectively, showing presence of colossal negative magnetoresistance in the system. The MR at 9 T for the $\mathbf{B}\perp c$ is significantly larger than for the $\mathbf{B} \parallel c$, indicating that the Fermi surface topology is anisotropic. Figure \ref{FigS16:FS} shows the calculated evolution of the Fermi surface at the $\Gamma$ point of the $p$-As from antiferromagnetic ($\theta$ = $\pi$/2) to ferromagnetic ($\theta$ = 0) spin alignment induced by the Eu-4\textit{f} moment at the Fermi level.

\begin{figure*}
\includegraphics[width=0.6\textwidth]{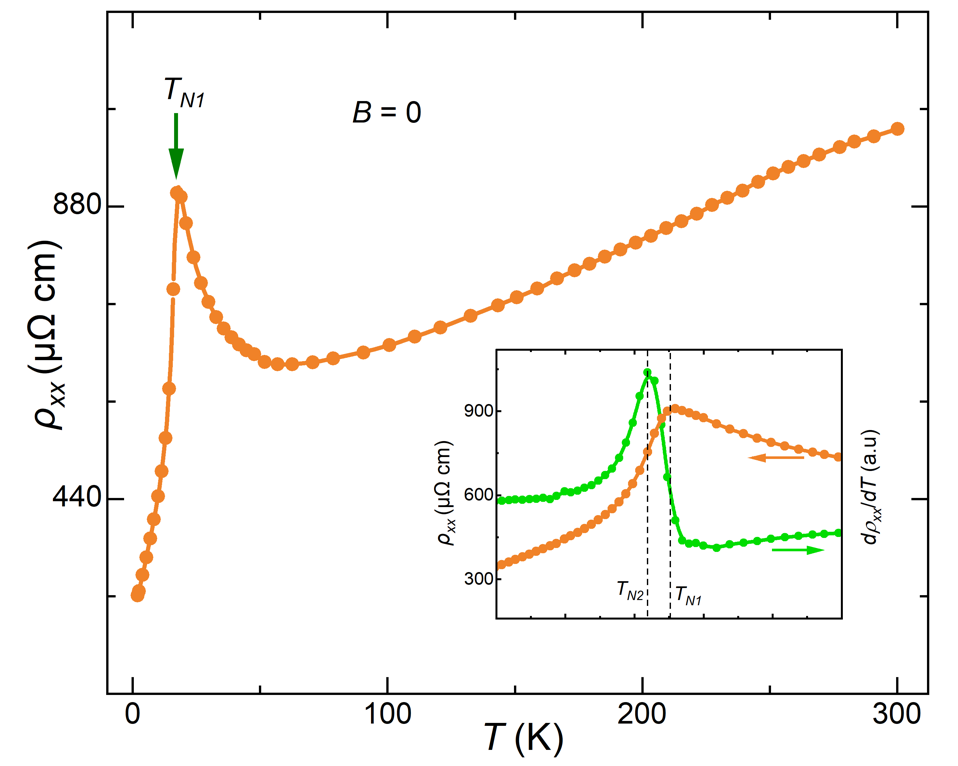}
\caption{Temperature-dependent resistivity, measured at 0T for $\mathbf{B}\parallel\textit{c}$. Inset: magnified area of resistivity near the transition temperature, as represented by the left axis.  The right axis represents the derivative of the same resistivity. $T_{N1}$ and $T_{N1}$ indicate the two transition temperatures compatible with the magnetization.}
\label{FigS4}
\end{figure*}

\begin{figure}
\includegraphics[width=\columnwidth]{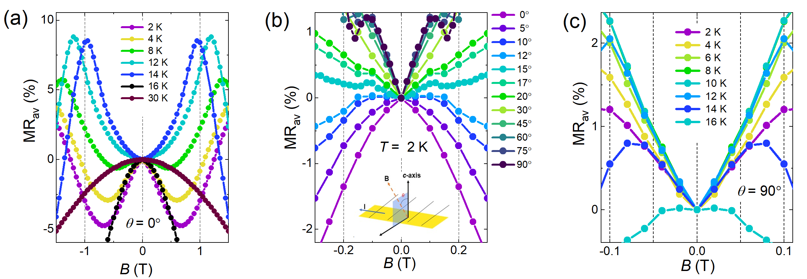}
\caption{(a) Field-dependent magnetoresistance (MR$_{av}$) recorded at various temperatures for 0$^\circ$ ($\mathbf{B}\parallel\textit{c}$). (b) MR$_{av}$ measured at various angles and at 2K. (c)  MR$_{av}$ recorded at different temperature and for 90$^\circ$ ($\mathbf{B}\perp\textit{c}$).} 
\label{FigS5}
\end{figure}

\begin{figure*}
\includegraphics[width=0.9\textwidth]{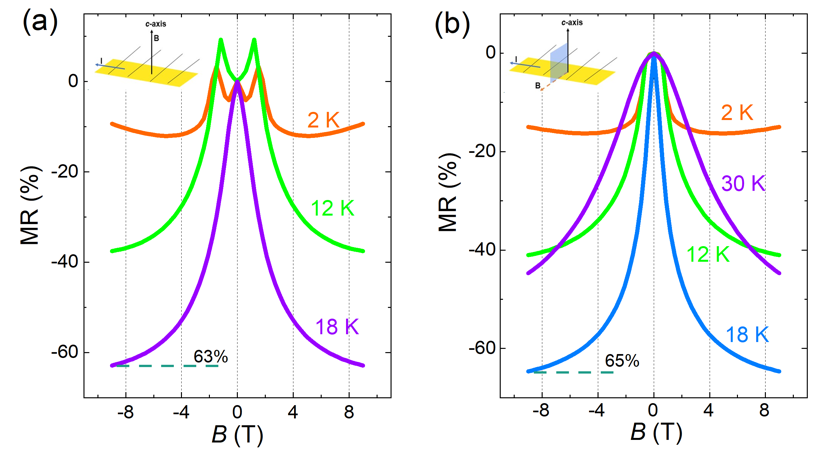}
\caption{Magnetoresistance recorded at different temperatures for (a) $\mathbf{B}\parallel\textit{c}$ (b) $\mathbf{B}\perp\textit{c}$.}
\label{FigS6}
\end{figure*}

\begin{figure}
\includegraphics[width=\columnwidth]{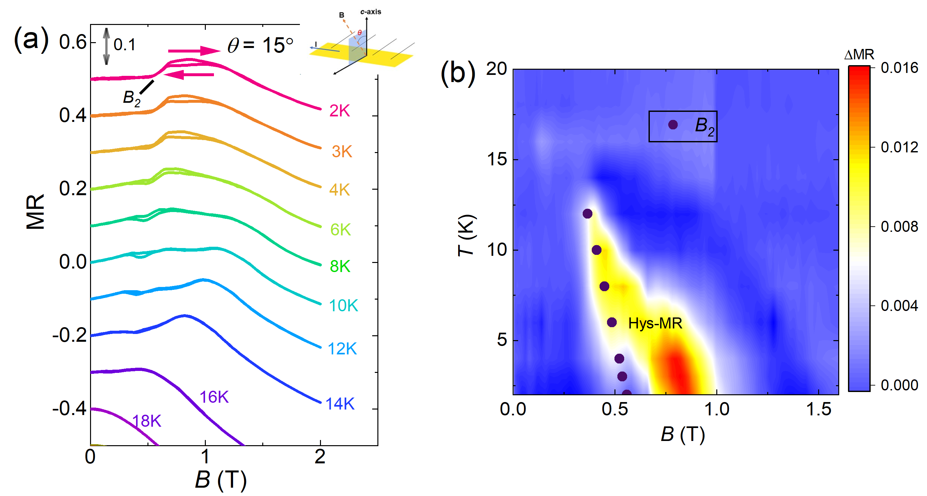}
\caption{(a) Magnetoresistance (MR) measured at different temperatures for $\theta = 15^\circ$ in swept-up and swept-down. This Figure is plotted in a shifted scale for clarity. (b) represents the color map of $\Delta$MR (=MR$^\mathrm{up}$-MR$^\mathrm{dn}$) calculated from data shown in (a). $B^\mathrm{up}_2$ (= $B_2$) represents the field strength for the transition in MR.} 
\label{15deg MR.png}
\end{figure}

\subsection*{3. Hall effect}

We have measured the $\rho_{yx}$ collected at various temperatures for $\theta$ = 0 (see Fig.~\ref{FigS8}a). The temperature dependence of carrier concentration is plotted in Fig.~\ref{FigS8}b. The plot in Fig.~\ref{FigS8}c illustrates the residual Hall effect ($\Delta\rho_{yx}$), for 0$^\circ$, displayed across the temperature range of 2-20 K. The extraction of $\Delta\rho_{yx}$ is performed using equation 1 in the high-field region and then extrapolated to the low-field region, as illustrated in Fig. \ref{figS9}. Below the $B_\mathrm{sat}$, a non-collinear magnetic structure arises with a maximum in the $\Delta\rho_{yx}$, which declines as temperatures increase. The maximum corresponds to the acquisition of finite spin chirality in the non-collinear magnetic structure \cite{Yan2022}. Furthermore, the $\Delta\rho_{yx}$ changes from positive to negative, as detailed in the main text, due to the competition of intrinsic and extrinsic transverse scattering or energy ($\omega$) modification.

\begin{figure*}
\includegraphics[width=0.6\textwidth]{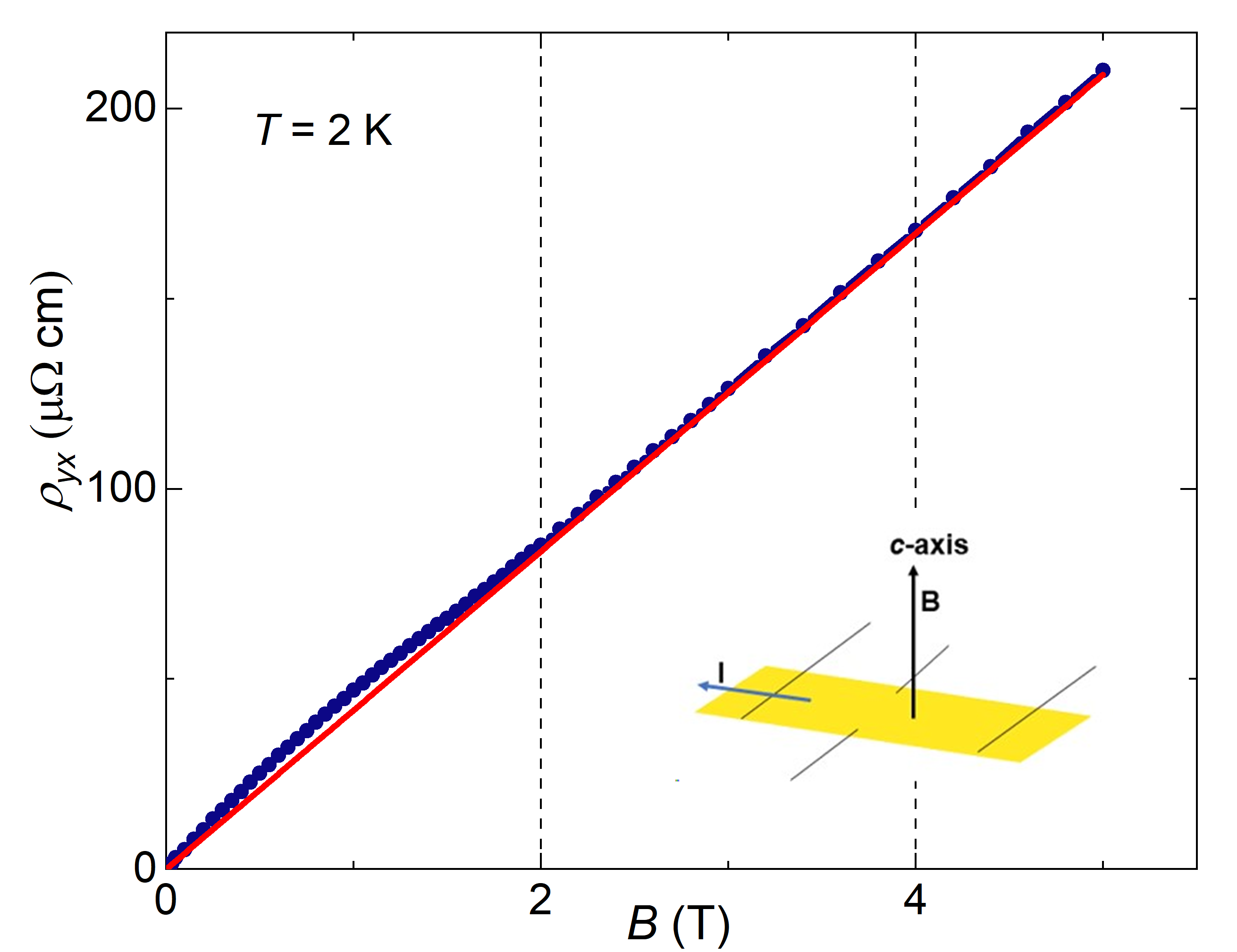}
\caption{Magnetic field dependent Hall resistivity at 2K. The red line represents the fit of the ordinary Hall effect (described in equation 1) to extract carrier concentrations. A similar procedure was used to extract the carrier concentration (shown in the inset of Figs. 4b and \ref{FigS8}(b)) for other temperatures and angles.}
\label{figS9}
\end{figure*}

\begin{figure*}
\includegraphics[width=0.8\textwidth]{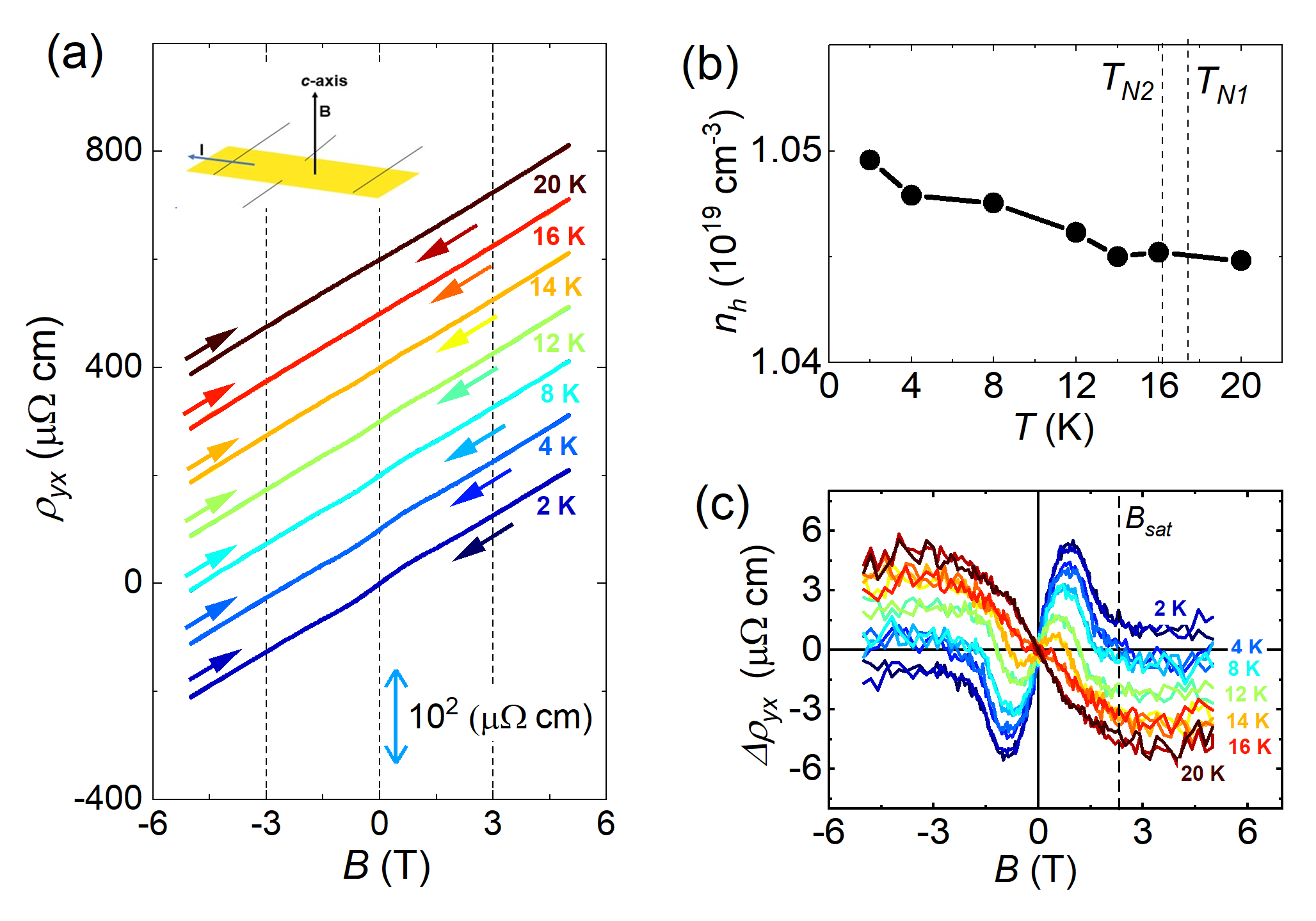}
\caption{(a) Hall resistivity as a function of magnetic field recorded at various temperatures for  $\mathbf{B} \parallel c$. The arrows represent the up and down directions of the field sweep. Data are offset for clarity. (b) Temperature dependence of carrier concentration. The dotted lines show the AFM transitions. (c) Anomalous Hall resistivity (obtained as described in the main text) as a function of the magnetic field at different temperatures. $B_\mathrm{sat}$ is the saturation field.} 
\label{FigS8}
\end{figure*}

\begin{figure}[h!]
\includegraphics[width=0.6\linewidth]{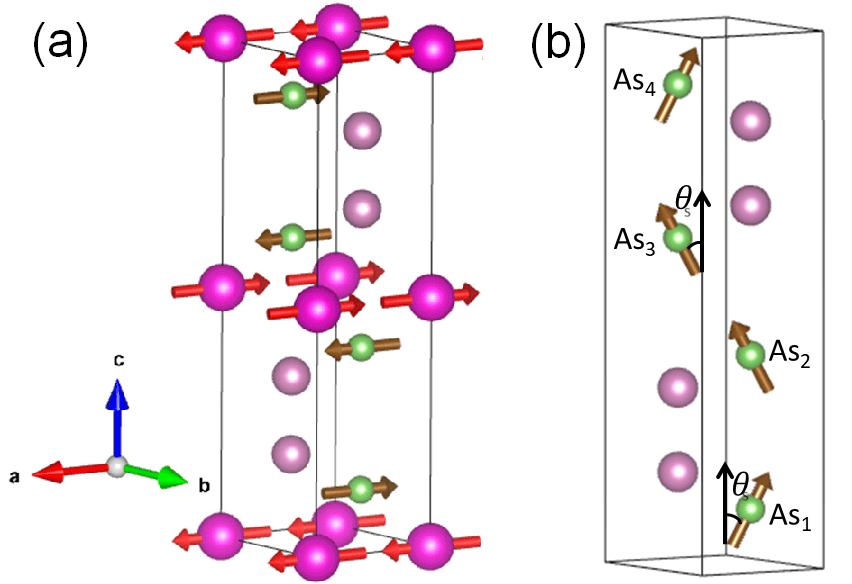}
\caption{(a) Crystal and magnetic structure of EuIn$_2$As$_2$. Regarding the crystal, the Eu atoms are purple, the As atoms are green and the In atoms are grey. Regarding the magnetism, we show the so-called A-type antiferromagnetic structure with the relevant spins on the Eu-\textit{f}- and As-\textit{p}-orbitals and with orientation along the \textit{a}-axis. The \textit{p}-orbitals of the As atoms present at the Fermi level have an induced magnetization antiparallel to that in the closest Eu layer; this induced magnetization is relevant for the AHE. The In atoms have no magnetization. (b) Flipping of the As-\textit{p} spins from the A-type antiferromagnetic structure with spin along the \textit{a}-axis to the ferromagnetic structure with spins along the \textit{c}-axis. We introduce the polar angle $\theta_S$ to simulate the rotation of the spins under an external magnetic field applied along the \textit{z}-axis. For $\theta_S=\pi$/2 we have the AFM configuration with spins along the \textit{a}-axis (AFMa), while for $\theta_S$=0 we have the FM phase with spin along the \textit{c}-axis (FMc). As$_1$ and As$_4$ have the spins of the first sublattice, while As$_2$ and As$_3$ have the spins of the second sublattice.}
\label{FigS12:Structure_1}
\end{figure}
\clearpage
\subsection*{4. Electronic and magnetic properties of the collinear magnetic phases}
Using the computational setup from the literature \cite{PhysRevB.108.075150} which combines the HSE approach and the Coulomb repulsion U, we compare the band inversion of the ferromagnetic phase with spins along the \textit{c}-axis (FMc) and altermagnetic phase with propagation vector $\tau_2$ = (0,0,1) and spins along the \textit{a}-axis [AMa, see Fig.~\ref{FigS12:Structure_1}(a)]. The system has band inversion at the $\Gamma$ point between \textit{s}-Cd and \textit{p}-As. The strength of the band inversion for both magnetic phases is shown in Fig.~\ref{FigS13:GAP_EuIn2As2} as a function of the Coulomb repulsion U. Within this approach, the experimental position of the \textit{f}-orbitals is reproduced for values of U between 1 and 2 eV \cite{PhysRevB.108.075150}. For these values of U, the system transits from an antiferromagnetic trivial insulator to a high-order topological ferromagnet known as ferromagnetic axion-insulator \cite{PhysRevMaterials.6.044204}.
Therefore, we can understand why we can expect a strong evolution of the electronic properties as a function of the magnetic field.

This $\tau_2$ = (0,0,1) phase is altermagnetic with bulk g-wave order as for MnTe \cite{LiborSmejkal2022}. The altermagnetic properties in the valence band are driven by the \textit{p}-orbitals of As since the Eu-4\textit{f}-states are not at the Fermi level and the Cd-\textit{s}-orbitals are nonmagnetic.
EuIn$_2$As$_2$ and MnTe have the same space group no. 194, however, in EuIn$_2$As$_2$, the Eu, In, and As atoms occupy Wyckoff 2a, 4f, and 4f positions, respectively. In MnTe, Mn and Te occupy 2a and 2c Wyckoff positions, respectively. Since we examined the magnetism coming from the As (with 4f Wyckoff position), the two compounds could have different properties for the Hall vector.
Depending on the symmetry, the anomalous Hall effect in altermagnets can also be present at zero field due to the breaking of the time-reversal symmetry. The direction of the Hall vector depends on the symmetries; additionally, a relativistic weak ferromagnetism appears along the Hall vector \cite{PhysRevB.103.L180407}. With the spins in the $ab$ plane but orthogonal to the lattice vectors, the Hall vector points along the \textit{c}-axis of MnTe \cite{PhysRevLett.132.176702,autieri2023dzyaloshinskiimoriya} due to a spin-orbit driven magnetic interaction which shares similarities with the Dzyaloshinskii–Moriya interaction. Among the first-neighbors magnetic atoms in 2a Wyckoff position with space group no. 194 as the Eu atoms in EuIn$_2$As$_2$, we have the same spin-orbit driven magnetic interaction that could produce weak ferromagnetism. However, this mechanism does not produce the non-collinearity experimentally observed; therefore, the origin of the non-collinearity in EuIn$_2$As$_2$ likely arises from the competition between different magnetic exchanges within the Eu-layers. 
\begin{figure}[h] 
\includegraphics[width=10.3cm,angle=0]{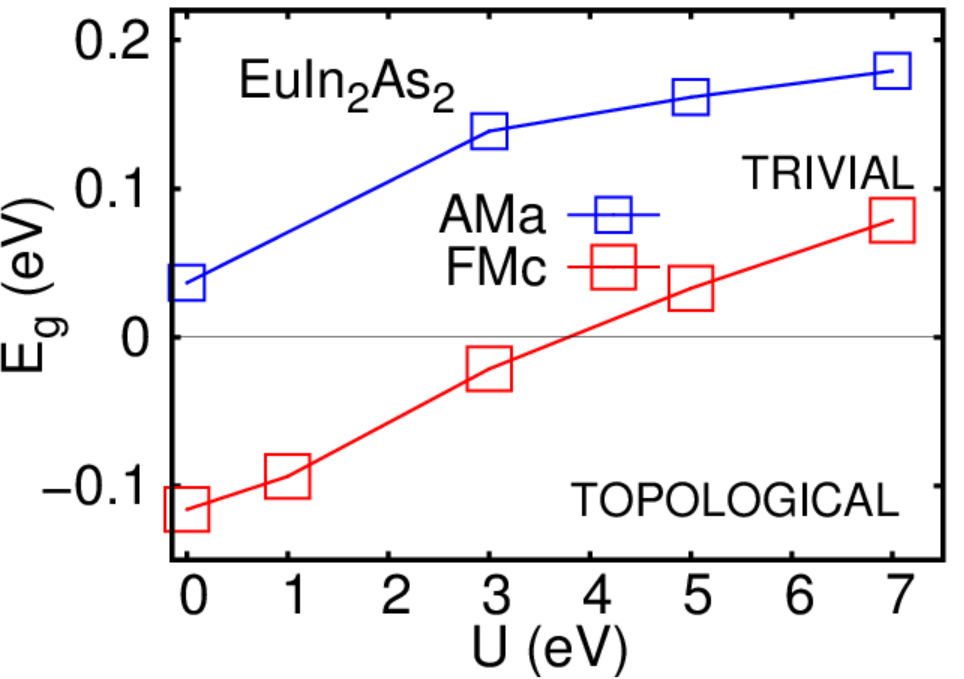}
\caption{Energy gap $E_g$ for EuIn$_2$As$_2$ as a function of the Coulomb repulsion energy $U$ in AFMa phase and in FMc phase. For the negative values of $E_g$ we are in the topological phase, while for positive values, we are in the trivial phase. These results were obtained using the previous computational setup reported in the literature \cite{PhysRevB.108.075150}.}
\label{FigS13:GAP_EuIn2As2}
\end{figure}

\begin{figure} 
    \subfloat[]{\includegraphics[width=0.4\textwidth]{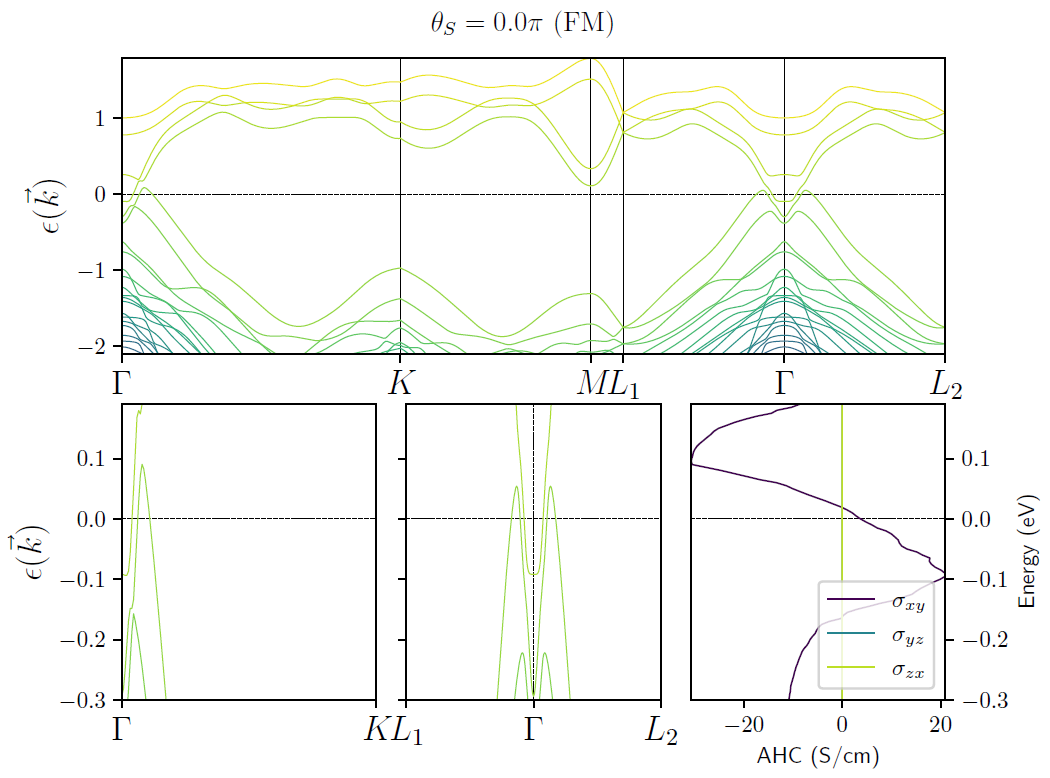}}
    \subfloat[]{\includegraphics[width=0.4\textwidth]{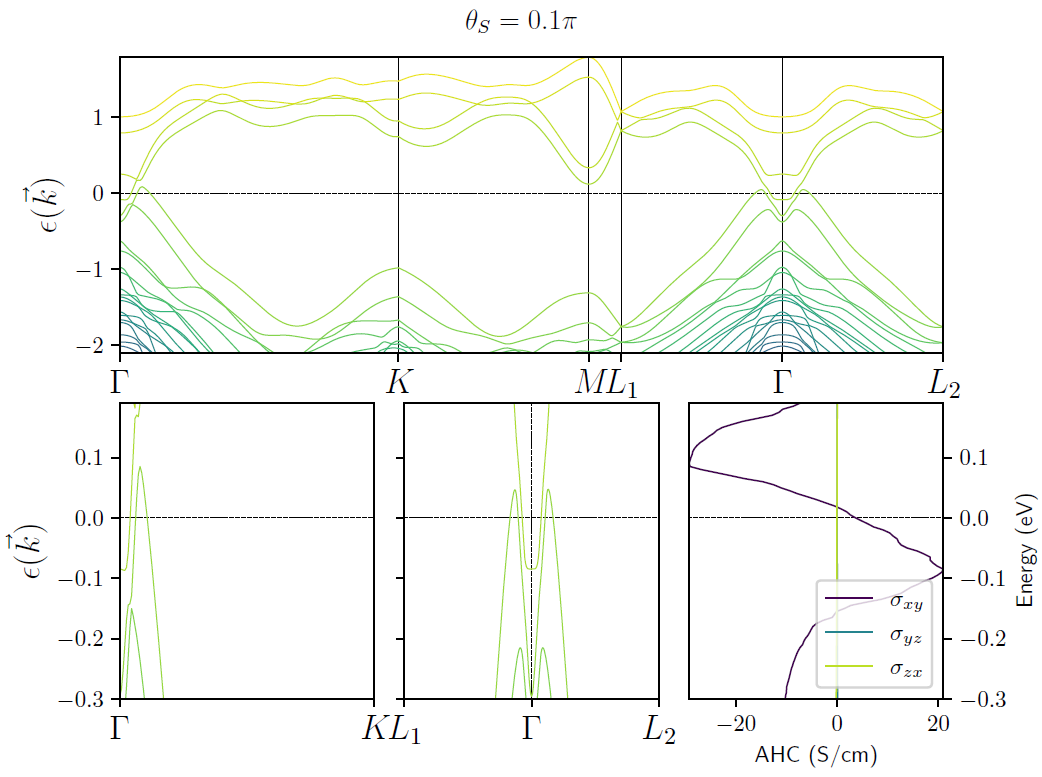}}\\
    \subfloat[]{\includegraphics[width=0.4\textwidth]{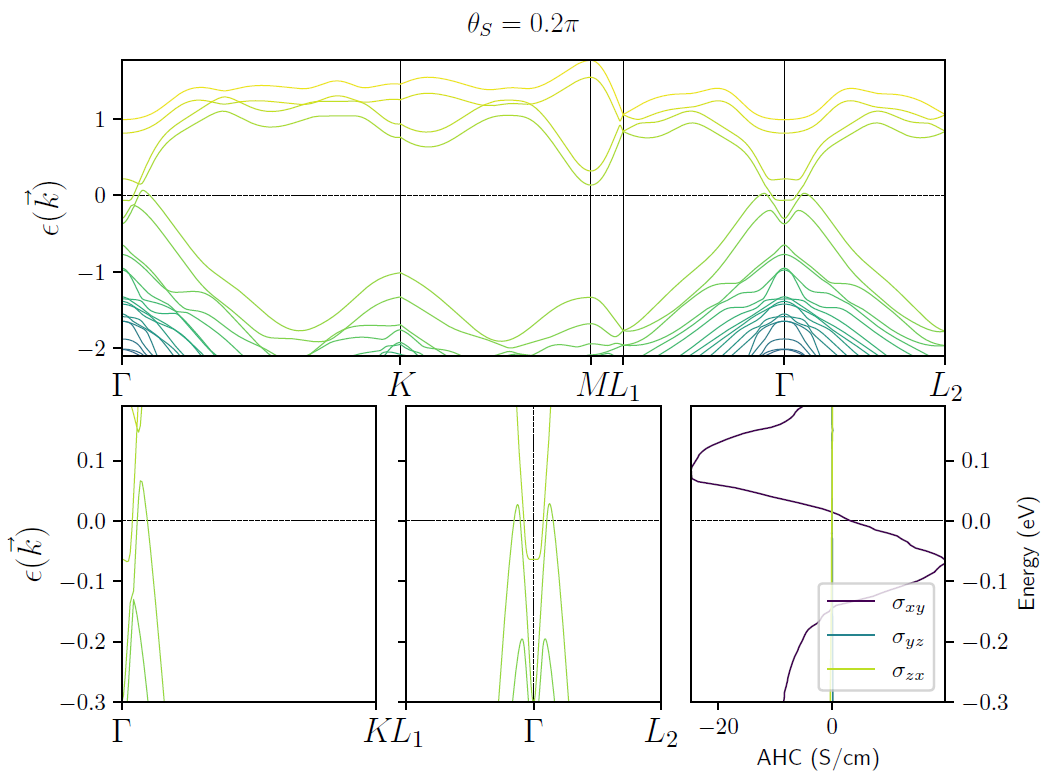}}
    \subfloat[]{\includegraphics[width=0.4\textwidth]{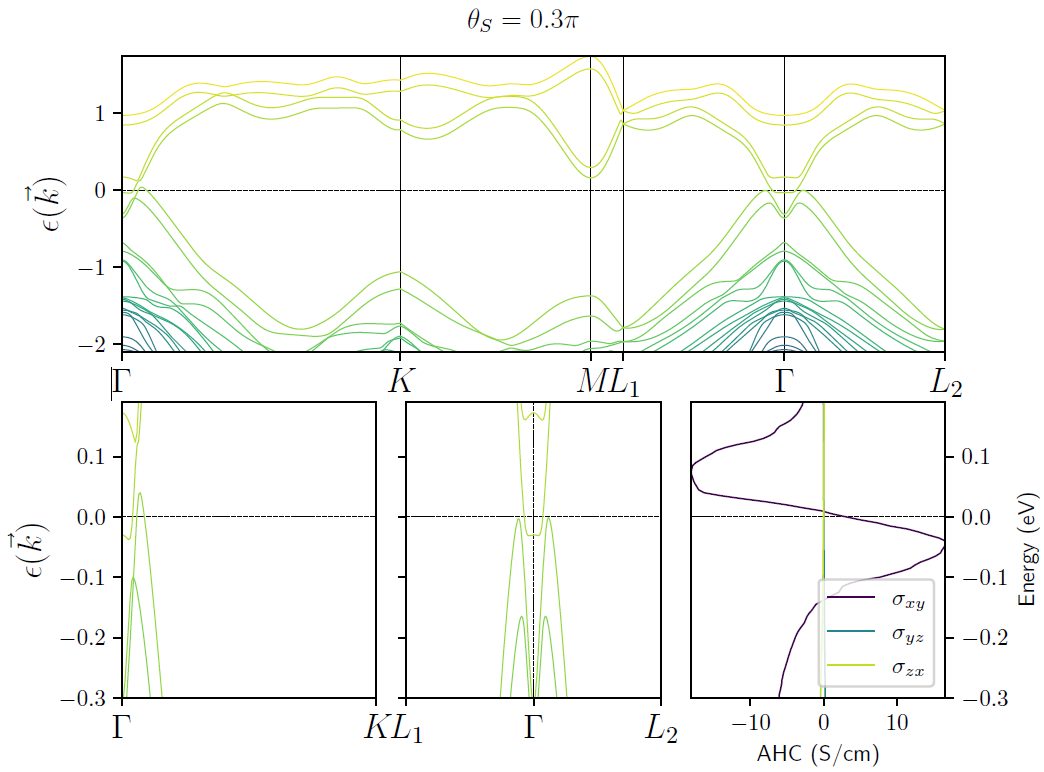}}\\
    \subfloat[]{\includegraphics[width=0.4\textwidth]{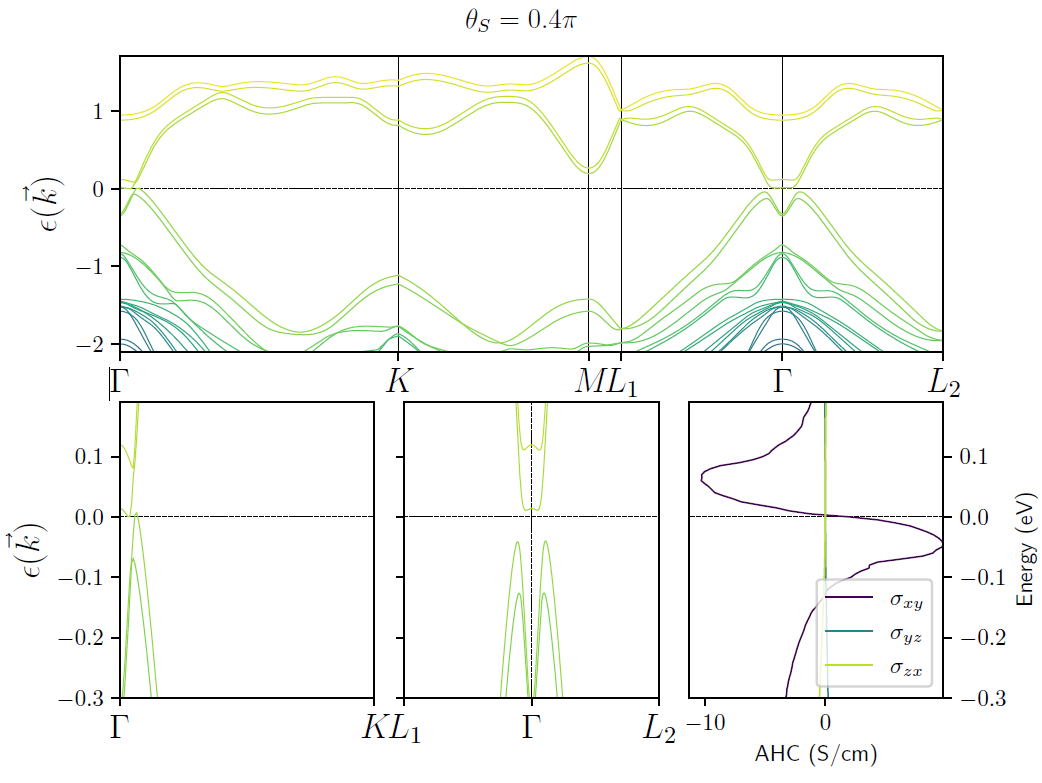}}
    \subfloat[]{\includegraphics[width=0.4\textwidth]{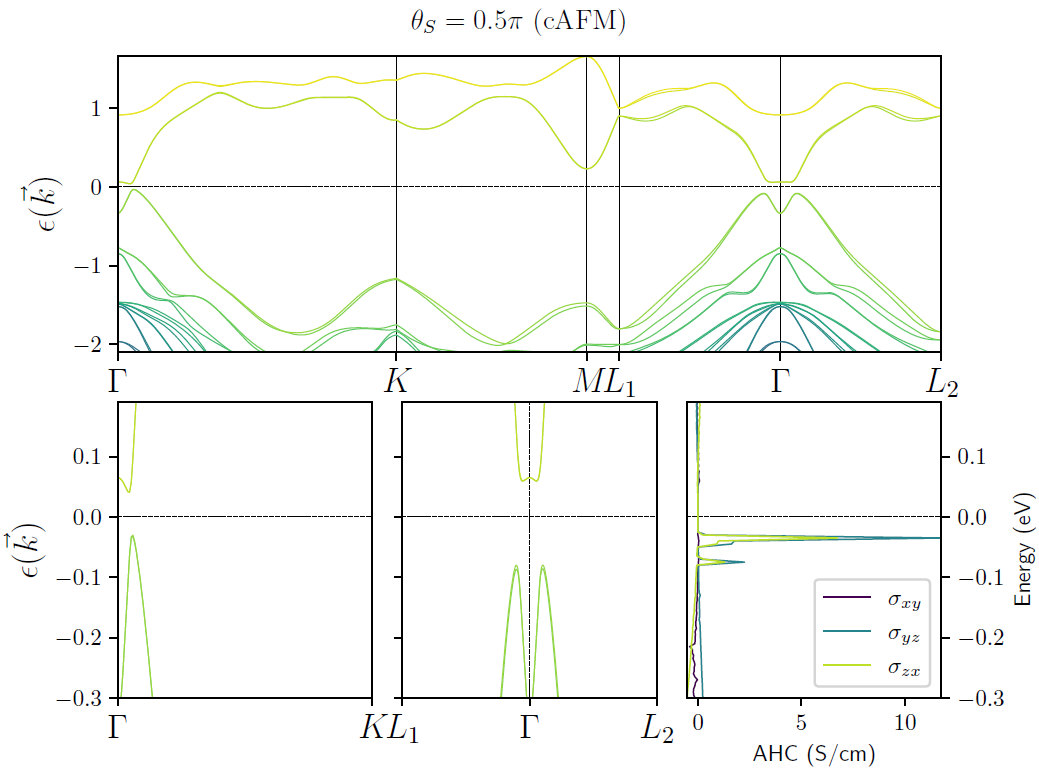}}
    \caption{Comparison between the dispersion (top) and the resulting intrinsic anomalous Hall conductivity (bottom right) for various $\theta_S$ angles between spin in As layers, depicted in Fig.~\ref{FigS12:Structure_1} (a). In the bottom left of the panels, we observe the low-energy part of the dispersion. Panels (a)-(f) show the evolution from the ferromagnetic arrangement ($\theta_S=0$) to the A-type antiferromagnet ($\theta_S=\pi$/2), in linear increments. The positive and negative peak in the valence and conduction bands derives from the properties of the Berry curvature for topological systems. Different colors are used for different energy bands as a guide for the eyes.}
    \label{FigS14:AHE_vs_theta}
\end{figure}

\begin{figure}[t] 
    \includegraphics[height=0.4\textheight]{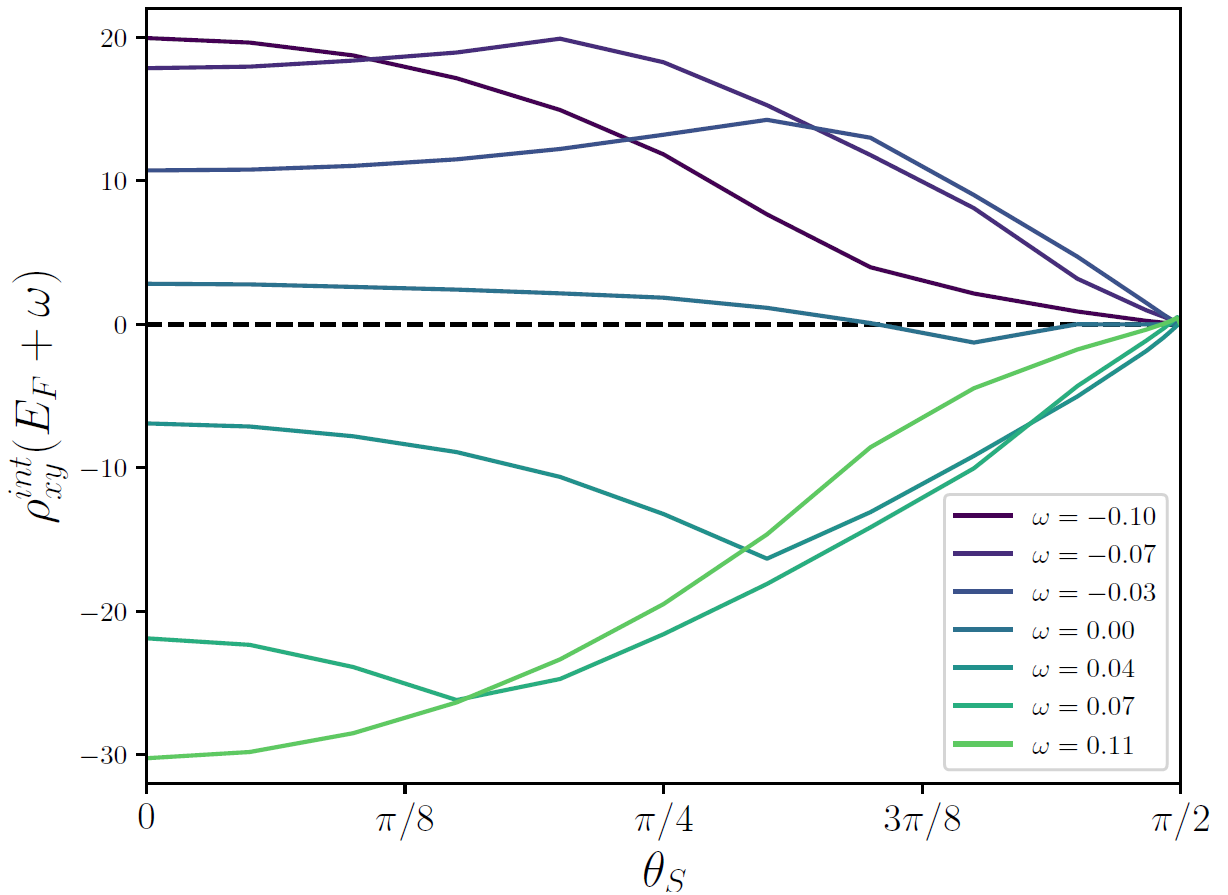}
    \caption{Evolution of the intrinsic anomalous Hall resistivity as a function of the $\theta_S$ polar angle for different energies $\omega$. The value of $\theta_S$=$\pi$ corresponds to zero magnetic field, while the value $\theta_S$=0 corresponds to the saturation magnetic field.}
    \label{FigS15:AHC_omega}
\end{figure}

\begin{figure}[b!] 
    \includegraphics[height=0.4\textheight]{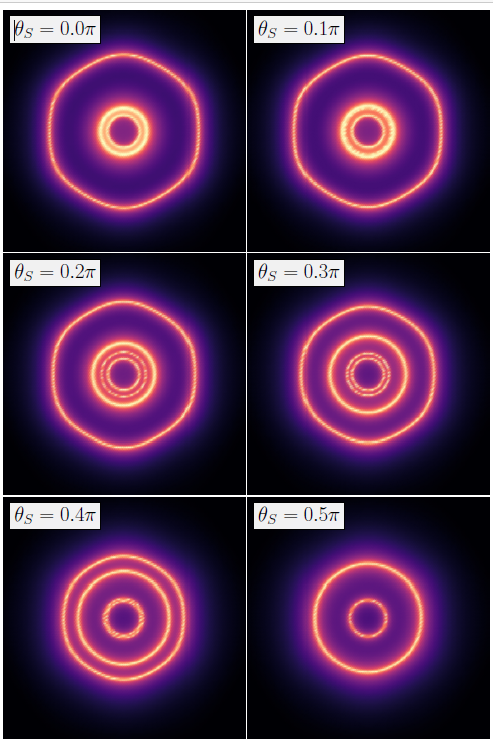}
    \caption{Evolution of the Fermi Surface around the $\Gamma$ point as spin alignment changes from FM with spins out of plane ($\theta_S=0$) to A-type AFM ($\theta_S=\frac{\pi}{2}$) for the energy $\omega$=-150 meV below the Fermi level expected for a stoichiometric compound.}
    \label{FigS16:FS}
\end{figure}
\subsection*{5. Evolution of the Fermi surface and band structure in applied magnetic fields combining first-principles and model Hamiltonian}

In the case of a perfectly stoichiometric compound, the system would be a trivial insulator or axion-insulator as discussed in the previous section. Experimentally, EuIn$_2$As$_2$ has Eu-deficiencies and this produces hole carriers with the Fermi level in the valence band, analogously to EuCd$_2$As$_2$ \cite{PhysRevLett.131.186704}. Due to the position of the Fermi level, the topological effects should not influence the experimentally observed electronic and magnetic properties. For the evolution of the Fermi surface and AHE, we study the electronic properties without including the electronic correlations; therefore, the band structure developed in this section would be in the topological phase. We study the evolution of the band structure as a function of the polar angle of the spin $\theta_S$ [see configuration in Fig.~\ref{FigS12:Structure_1}(b)] as it was analyzed in doped antiferromagnetic insulators \cite{chowdhury2022topological}.

For a realistic simulation of this compound, we need to have As$_1$ and As$_4$ belonging to one sublattice while As$_2$ and As$_3$ belong to the other sublattice, as shown in Fig.~\ref{FigS12:Structure_1}(b). 
We start from the non-magnetic (NM) Hamiltonian obtained with the Wannier functions basis:
\begin{equation}
H^{NM}= \sum_{i,j} t_{i,j}^{l,m}c^{\dag}_{i,l}c_{j,m}
\label{eqnS3}
\end{equation}
The non-magnetic Hamiltonian already contains the crystal symmetries of the NiAs space group. Once the local spin-splitting is added, the altermagnetic spin-splitting will appear. We add the on-site terms as the on-site spin-splitting $\Vec{h}(\theta_S,\phi_S)$ dependent on the angles and the 
As spin-orbit coupling $\lambda_{As}$=164 meV \cite{Wadge_2022}.
\begin{equation}
H^{split}+H^{SOC}= \sum_i(-\Vec{h}(\theta_S,\phi_S)_i\cdot\Vec{S}_i+\lambda_i\Vec{L}_i\cdot\Vec{S}_i) 
\label{eqnS4}
\end{equation}
while for the \textit{s}-orbitals there is no SOC. The \textit{s}-orbitals have neither spin-splitting, while the module of the on-site spin-splitting of As is fixed at $|h|$=300 meV. The crucial information regarding the spin evolution is included in the real space magnetic configuration $\Vec{S}_i$ and their angular dependence of the As spin described in Fig.~\ref{FigS12:Structure_1}(b). 
In the general case, the intrinsic anomalous Hall effect $\sigma^{int}_{xy}$ in an altermagnet with flipping spins could be seen as a combination of the anomalous Hall effect from the altermagnet (which depends on the S$_x$ component) and the anomalous Hall effect from the ferromagnetism along the \textit{c}-axis (which depends on the S$_z$ component). For $\theta_S$=$\frac{\pi}{2}$, the AHE comes only from the altermagnetic properties, while for $\theta_S$=0 the AHE depends only on the properties of the ferromagnet. For intermediate values of $\theta_S$, the interplay is not straightforward since we cannot decompose the AHE calculated in the \textbf{k}-space in terms of the real space quantities as the polar angle $\theta_S$, a similar aspect was proven for a decomposition of the AHE on different layers (see supp. material of the reference \cite{PhysRevLett.127.127202}).

We study the AHE with spin along the \textit{c}-axis in the topological regime since we are interested in matching the experimental results in the limit of B$>B_{sat}$ where the system becomes a ferromagnetic axion insulator \cite{PhysRevMaterials.6.044204}.
We calculate the AHE as a function of $\theta_S$, where in the experimental realization, the polar angle is controlled by the magnetic field. The results of the AHC are reported in Fig.~\ref{FigS14:AHE_vs_theta}. Additionally, we calculate the $\sigma^{int}_{xy}$ as a function of $\theta_S$ for different energies $\omega$ and the results are shown in Fig.~\ref{FigS15:AHC_omega} where we also observe a positive sign of AHC for the three lowest values of $\omega$, while it is negative for the four highest values with a sign change of the AHC.
For negative values of $\omega$ the Fermi level lies in the valence band and the anomalous Hall conductivity gets reduced in an applied magnetic field in qualitative agreement with the experimental results.

We also calculated the evolution of the Fermi surface topology at the energy $\omega$=-150 meV below the Fermi level for a stoichiometric compound. We obtain a strong evolution of the Fermi surfaces from the AMa to the FMc phase. The Fermi is calculated at k$_z$=0, which is a nodal plane for the altermagnetic phase. In the collinear AMa phases, we can see 2 couples of Fermi surfaces since the altermagnetic spin-splitting is absent at k$_z$=0. When we include an S$_Z$ component, we have transitioned to a nodeless band structure \cite{PhysRevB.109.024404} with 4 distinct Fermi sheets as shown in Fig.~\ref{FigS16:FS}, which are 2 for spin-up and 2 for spin-down. Finally, for the FM phase and close to it ($\theta_S$=0, 0.1$\pi$), we are approaching a Lifshitz transition but still have 4 Fermi surfaces. Therefore, the system is prone to have Lifshitz phase transitions induced by the external magnetic field along the \textit{z}-axis; the Lifshitz phase transition depends on tiny details such as small variations in the filling or the orientation of the magnetic field. 
However, the in-plane magnetic field produced the Lifshitz phase transition. The magnetic-field induced Lifshitz phase transitions are usually concomitant to metamagnetic transitions \cite{doi:10.1126/sciadv.abo7757}, especially for compounds hosting f-electrons \cite{PhysRevB.90.075127,PhysRevB.86.075108,Shivaram2019,McCollam_2021}. Therefore, the metamagnetism experimentally observed can be generated by the Lifshitz phase transition with the in-plane magnetic field. 



\end{document}